\documentstyle{article}

\input epsf

\newcommand{\bea}{\begin{eqnarray}}
\newcommand{\eea}{\end{eqnarray}}

\begin{document}

%%%%%%%%%%%%%%%%%%%%%%%%%%%%%%%%%%%%%%%%%%%%%%%%%%%%%%%%%%%%%%%
%\draft
%  For 2 column format.
%@twocolumnfalse\endcsname

%%%%%%%%%%%%%%%%%%%%%%%%%%%%%%%%%%%%%%%%%%%%%%%%%%%%%%%%%%%%%%%%%
\title{Cosmological Perturbations with Multiple Fluids and Fields}
\author{Jai-chan Hwang${}^{(a,b)}$ and Hyerim Noh${}^{(c,b)}$ \\
        \footnotesize
        ${}^{(a)}$ Department of Astronomy and Atmospheric Sciences, 
                   Kyungpook National University, Taegu, Korea \\
        \footnotesize
        ${}^{(b)}$ Institute of Astronomy, Madingley Road, Cambridge, UK \\
        \footnotesize
        ${}^{(c)}$ Korea Astronomy Observatory, Taejon, Korea}
\date{\today}
\maketitle
%\footnote{jchan@knu.ac.kr} 
%\footnote{hr@kao.re.kr} 

%%%%%%%%%%%%%%%%%%%%%%%%%%%%%%%%%%%%%%%%%%%%%%%%%%%%%%%%%%%%%%%
\begin{abstract}

We consider the evolution of perturbed cosmological spacetime with multiple
fluids and fields in Einstein gravity.
Equations are presented in gauge-ready forms, and are presented
in various forms using the curvature ($\Phi$ or $\varphi_\chi$) 
and isocurvature ($S_{(ij)}$ or $\delta \phi_{(ij)}$) perturbation variables
in the general background with $K$ and $\Lambda$.
We clarify the conditions for conserved curvature and isocurvature
perturbations in the large-scale limit.
Evolutions of curvature perturbations in many different gauge conditions 
are analysed extensively.
In the multi-field system we present a general solution 
to the linear order in slow-roll parameters.

\end{abstract}
%%%%%%%%%%%%%%%%%%%%%%%%%%%%%%%%%%%%%%%%%%%%%%%%%%%%%%%%%%%%%%%%%

\noindent
\hskip 1cm
PACS numbers: 98.80.Cq, 98.80.Hw, 98.70.Vc, 04.62+v

%%%%%%%%%%%%%%%%%%%%%%%%%%%%%%%%%%%%%%%%%%%%%%%%%%%%%%%%%%%%%%%
%  For 2 column format.
%\vskip2pc]
%%%%%%%%%%%%%%%%%%%%%%%%%%%%%%%%%%%%%%%%%%%%%%%%%%%%%%%%%%%%%%%

%%%%%%%%%%%%%%%%%%%%%%%%%%%%%%%%%%%%%%%%%%%%%%%%%%%%%%%%%%%%%%%%%
\section{Introduction}
                                          \label{sec:Introduction}

Cosmological models with multiple components of underlying energy-momentum
content are not only general but also necessarily appear in many
cosmological situations including the early and the late evolution stages
of our world models. 
Assuming the gauge degrees of freedom are properly fixed, for 
the $n$ component medium with fluids or fields in Einstein gravity
we anticipate a set of $n$ coupled second-order
differential equations for the scalar-type perturbation,
a set of $n$ (in general) coupled first-order differential equations
for the vector-type perturbation, and one second-order differential
equation for the tensor-type perturbation.
The situations of the vector- and tensor-type perturbations are rather
trivial to handle, and, in fact, the situation is similar even in some classes 
of generalized versions of gravity theories 
\cite{Hwang-1991-PRW,Hwang-Noh-2001-CMBR}. 
However, the scalar-type perturbation in such a multi-component system is
naturally more complicated, and often requires numerical methods.
Not only the equations are complicated for the scalar-type
perturbation, the behavior and the way of handling the equations
depend on our choice of the gauge condition which we need to make.
In either analytic or numerical handling of the scalar-type perturbations, 
it is useful and is often necessary to examine the problems from the 
perspectives of different gauge conditions.

In the following, for the benefit of future studies,
we would like to present various useful forms of
the scalar-type perturbation equations considering arbitrary numbers
of imperfect fluids and scalar fields with general mutual interactions
among the components in Einstein gravity.
Equations are presented in the gauge-ready forms which allow to take
any of our favorite gauge conditions easily, and also allow easy 
translation of a solution derived in one gauge condition to the ones 
in other gauge conditions.
We also present closed form equations in terms of the curvature and
isocurvature perturbation variables in several useful forms, 
and derive general solutions if available.
We recently have presented similar work on systems with multiple minimally 
coupled scalar fields \cite{Hwang-Noh-2000-MSFs}.
In the present work we present a complementary part now
including the multiple fluids as well as the interactions among
fluids and fields; we also extend \cite{Hwang-Noh-2000-MSFs} 
by considering general $K$ and $\Lambda$ 
and the fluid formulation of the field perturbations in \S \ref{sec:MSFs}. 

The manuscript is mainly concerned with the formal aspects of perturbations
in the multi-component situations.
There are some advantages of investigating 
perturbations in a formal way keeping their most general forms.
{}For example, we emphasize that the basic set of equations
for a single component case remains valid even in the multi-component
situations with additional equations for the components.
Also, by including most general energy-momentum tensor in the
fluids system we will show that the same equations remain valid
for the mixture of fluids and fields, and sometimes even for the
generalized gravity theories.
By comparing the fluid system and field system we can clarify the
similarity and difference between them.
Although, we intentionally have kept the presentation formal, many new 
results are presented together with the known ones;
see \S \ref{sec:Summary} for a summary.
Previous studies of perturbations in the multi-component fluid system 
in a similar spirit can be found in 
\cite{Kodama-Sasaki-1984,Kodama-Sasaki-1986,Kodama-Sasaki-1987,multi,%
Hwang-1991-PRW,Hwang-Noh-2001-CMBR}.
We start by introducing our notations, fundamental equations, and strategy
in \S \ref{sec:basic}.
We set $c \equiv 1$.

%%%%%%%%%%%%%%%%%%%%%%%%%%%%%%%%%%%%%%%%%%%%%%%%%%%%%%%%%%%%%%%%%
\section{Perturbed world model and basic equations}
                                          \label{sec:basic}

We consider an arbitrary number of mutually interacting imperfect fluids 
in Einstein gravity. 
The gravitational field equation and the energy-momentum conservation
equation are
\bea
   & & G_{ab} \equiv 8 \pi G T_{ab} - \Lambda g_{ab}, \quad T^b_{a;b} = 0.
   \label{GFE}
\eea
The energy-momentum tensor can be decomposed into the sum of the
individual one as
\bea
   & & T_{ab} = \sum_{l} T_{(l)ab},
   \label{Tab-sum}
\eea
and the energy-momentum conservation gives
\bea
   & & T^{\;\;\;\; b}_{(i)a;b} \equiv Q_{(i)a}, \quad
       \sum_{l} Q_{(l)a} = 0,
   \label{Tab-i-conservation}
\eea
where $(i)$ indicates the $i$-th component of $n$ matters (including
fluids and fields) with $i,j,k, \dots = 1,2,\dots, n$.

We consider the general perturbations in the FLRW world model.
In the spatially homogeneous and isotropic background we are considering,
to the linear order the three different types of perturbation
decouple from each other and evolve independently.
The presence of tensor-type anisotropic stress contributes as the source or 
sink of the graviational wave of the perturbed spacetime. 
The rotational perturbation of the individual fluid is described by 
the angular-momentum conservation where the vector-type anisotropic
stress and the mutual interaction terms among fluids can work as the
source or sink of the rotation of individual component.
The most general situations of vector- and tensor-type perturbations
are presented in \cite{Hwang-1991-PRW,Hwang-Noh-2001-CMBR}.
In the following we will consider the {\it scalar-type} perturbation only.

A metric with the most general scalar-type perturbation in the FLRW 
background is
\bea
   & & d s^2 = - a^2 \left( 1 + 2 \alpha \right) d \eta^2 
       - 2 a^2 \beta_{,\alpha} d \eta d x^\alpha
       + a^2 \left[ g^{(3)}_{\alpha\beta} \left( 1 + 2 \varphi \right)
       + 2 \gamma_{,\alpha|\beta} \right] d x^\alpha d x^\beta,
   \label{metric-general}
\eea
where $a(\eta)$ is the cosmic scale factor.
$\alpha$, $\beta$, $\gamma$ and $\varphi$
are spacetime dependent perturbed order variables.
A vertical bar $|$ indicates a covariant derivative based on 
$g^{(3)}_{\alpha\beta}$.
Considering the FLRW background and the scalar-type perturbation,
the energy-momentum tensor can be decomposed into the collective fluid 
quantities as
\bea
   & & T^0_0 = - \left( \bar \mu + \delta \mu \right), \quad
       T^0_\alpha = - {1 \over k} \left( \mu + p \right) v_{,\alpha}, \quad
       T^\alpha_\beta = \left( \bar p + \delta p \right) \delta^\alpha_\beta
       + \left( {1 \over k^2} \nabla^\alpha \nabla_\beta 
       + {1 \over 3} \delta^\alpha_\beta \right) \pi^{(s)},
   \label{Tab}
\eea
where $\mu (\equiv \bar \mu + \delta \mu)$, 
$p (\equiv \bar p + \delta p)$, $v$, and $\pi^{(s)}$ are
the energy density, the isotropic pressure, the velocity or
flux related variable, and the anisotropic stress, respectively;
$k$ is the comoving wavenumber.
In terms of the individual matter's fluid quantities we have
\bea
   & & \bar \mu = \sum_{l} \bar \mu_{(l)}, \quad
       \delta \mu = \sum_{l} \delta \mu_{(l)},
   \label{fluid-sum}
\eea
and similarly for $\bar p$, $\delta p$, $(\mu + p) v$, and $\pi^{(s)}$;
an overbar indicates the background order quantity, and will be ignored
unless necessary.
The interaction terms among fluids introduced in
eq. (\ref{Tab-i-conservation}) are decomposed as
\bea
   & & Q_{(i)0} \equiv - a \left[ \bar Q_{(i)} ( 1 + \alpha )
       + \delta Q_{(i)} \right], \quad
       Q_{(i)\alpha} \equiv J_{(i),\alpha}.
   \label{Q-decomposition}
\eea

The equations for the background are:
\bea
   & & H^2 = {8 \pi G \over 3} \mu + {\Lambda \over 3} - {K \over a^2}, \quad
       \dot H = - 4 \pi G (\mu + p) + {K \over a^2}, 
   \label{BG1} \\
   & & \dot \mu_{(i)} + 3 H \left( \mu_{(i)} + p_{(i)} \right) = Q_{(i)},
   \label{BG2}
\eea
where an overdot indicates the time derivative based on $t$
defined as $dt \equiv a d \eta$.
$K$ is the normalized spatial curvature, and $H \equiv \dot a/a$.
The two equations in eq. (\ref{BG1}) follow from $G^0_0$ and 
$G^\alpha_\alpha - 3 G^0_0$ components of the field equation in 
eq. (\ref{GFE}), respectively,
and eq. (\ref{BG2}) follows from eq. (\ref{Tab-i-conservation}).
By adding eq. (\ref{BG2}) over components we have
$\dot \mu + 3 H ( \mu + p ) = 0$ which also follows from eq. (\ref{BG1}).

We present a complete set of equations describing the scalar-type 
perturbation without fixing the temporal gauge condition, i.e., in a 
gauge-ready form \cite{Bardeen-1988,Hwang-1991-PRW,Hwang-Noh-2001-CMBR}:
\bea
   & & \kappa \equiv 3 \left( - \dot \varphi + H \alpha \right)
          + {k^2 \over a^2} \chi,
   \label{G1} \\
   & & - {k^2 - 3K \over a^2} \varphi + H \kappa
       = - 4 \pi G \delta \mu,
   \label{G2} \\
   & & \kappa - {k^2 - 3 K \over a^2} \chi
       = 12 \pi G {a \over k} ( \mu + p ) v,
   \label{G3} \\
   & & \dot \chi + H \chi - \alpha - \varphi 
       = 8 \pi G {a^2 \over k^2} \pi^{(s)},
   \label{G4} \\
   & & \dot \kappa + 2 H \kappa
       + \left( 3 \dot H
       - {k^2 \over a^2} \right) \alpha
       = 4 \pi G \left( \delta \mu + 3 \delta p \right),
   \label{G5} \\
   & & \delta \dot \mu_{(i)} + 3 H \left( \delta \mu_{(i)}
       + \delta p_{(i)} \right)
       = - {k \over a} \left( \mu_{(i)} + p_{(i)} \right) v_{(i)}
       + \dot \mu_{(i)} \alpha
       + \left( \mu_{(i)} + p_{(i)} \right) \kappa + \delta Q_{(i)}, 
   \label{G6} \\
   & & {1 \over a^4 ( \mu_{(i)} + p_{(i)} )}
       \left[ a^4 ( \mu_{(i)} + p_{(i)} ) v_{(i)} \right]^\cdot
       = {k \over a} \Bigg[ \alpha
       + {1 \over \mu_{(i)} + p_{(i)}} \left( \delta p_{(i)}
       - {2 \over 3} {k^2 - 3 K \over k^2} \pi^{(s)}_{(i)} - {J}_{(i)} \right)
       \Bigg],
   \label{G7}
\eea
where $\chi \equiv a ( \beta + a \dot \gamma )$.
$\chi$, $\varphi$ and $\kappa$ correspond to the shear,
the three-space curvature and the perturbed expansion
of the normal-frame vector field, respectively, see
\cite{Hwang-1991-PRW,Hwang-Noh-2001-CMBR} for details.
$\nabla_\alpha$ and $\Delta$ are the covariant derivative and 
the Laplacian based on $g_{\alpha\beta}^{(3)}$.
Equations (\ref{G1}-\ref{G5}) are:
the definition of $\kappa$, 
ADM energy constraint ($G^0_0$ component of the field equation),
momentum constraint ($G^0_\alpha$ component), 
tracefree part of ADM propagation
($G^\alpha_\beta - {1 \over 3} \delta^\alpha_\beta G^\gamma_\gamma$ component),
and Raychaudhuri equation ($G^\gamma_\gamma - G^0_0$ component), respectively.
Equations (\ref{G6},\ref{G7}) follow from the energy conservation 
($T_{(i)0;b}^{\;\;\;\; b} = Q_{(i)0}$) and the momentum conservation 
($T_{(i)\alpha;b}^{\;\;\;\; b} = Q_{(i)\alpha}$), respectively.
We note that eqs. (\ref{G6},\ref{G7}) in the same forms are valid even in 
a class of generalized gravity theories
considered in \cite{Hwang-Noh-2001-CMBR}; 
see eqs. (46,47) in \cite{Hwang-Noh-2001-CMBR}.
Equations (\ref{G1}-\ref{G5}) also remain valid in the generalized 
gravity where we can reinterprete the fluid quantities as the effective ones;
see the Appendix B of \cite{Hwang-Noh-2001-CMBR}.

By adding properly eqs. (\ref{G6},\ref{G7}) over all components of the fluids,
and using properties in eq. (\ref{fluid-sum}),
we get equations for the collective fluid quantities as:
\bea
   & & \delta \dot \mu + 3 H \left( \delta \mu + \delta p \right)
       = ( \mu + p ) \left( \kappa - 3 H \alpha - {k \over a} v \right),
   \label{G8} \\
   & & {\left[ a^4 ( \mu + p ) v \right]^\cdot \over a^4 ( \mu + p )} 
       = {k \over a} \left[ \alpha + {1 \over \mu + p} \left( \delta p
       - {2 \over 3} {k^2 - 3 K \over k^2} \pi^{(s)} \right) \right].
   \label{G9}
\eea
We introduce
\bea
   & & e_{(i)} \equiv \delta p_{(i)} - c_{(i)}^2 \delta \mu_{(i)}, \quad
       c_{(i)}^2 \equiv {\dot p_{(i)} / \dot \mu_{(i)}}, \quad
       e \equiv \delta p - c_s^2 \delta \mu, \quad
       c_s^2 \equiv {\dot p / \dot \mu},
   \nonumber \\
   & & w_{(i)} \equiv p_{(i)} / \mu_{(i)}, \quad
       w \equiv p/\mu,
   \label{e-def}
\eea
where $e_{(i)}$ and $e$ are the entropic perturbations.
Equation (\ref{G6}) can be written in another form as
\bea
   & & \dot \delta_{(i)}
       + 3 H \left( c_{(i)}^2 - w_{(i)} \right) \delta_{(i)}
       + 3 H \left( 1 + w_{(i)} \right) q_{(i)} \delta_{(i)}
   \nonumber \\
   & & \qquad
       = \left( 1 + w_{(i)} \right) \left[ \kappa 
       - 3 H \left( 1 - q_{(i)} \right) \alpha - {k \over a} v_{(i)} \right]
       + {1 \over \mu_{(i)}} \left( - 3 H e_{(i)} + \delta Q_{(i)} \right),
   \label{G6-1}
\eea
where $Q_{(i)} \equiv 3 H ( \mu_{(i)} + p_{(i)} ) q_{(i)}$, and
we used
$\dot w_{(i)} = - 3 H ( c_{(i)}^2 - w_{(i)} ) ( 1 + w_{(i)} ) ( 1 - q_{(i)} )$;
$\delta_{(i)} \equiv \delta \mu_{(i)}/\mu_{(i)}$.

In the following we briefly summarize the gauge-ready strategy suggested 
by Bardeen \cite{Bardeen-1988} and elaborated in \cite{Hwang-1991-PRW};
see \cite{Hwang-Noh-2001-CMBR} for a recent extension.
Under the gauge transformation $\tilde x^a = x^a + \xi^a$
with $\xi^t \equiv a \xi^0$ ($0 = \eta$) the perturbed
metric and fluid quantities change as 
\cite{Hwang-1991-PRW,Hwang-Noh-2001-CMBR}:
\bea
   & & \tilde \alpha = \alpha - \dot \xi^t, \quad
       \tilde \varphi = \varphi - H \xi^t, \quad
       \tilde \kappa = \kappa + \left( 3 \dot H + {\Delta \over a^2}
       \right) \xi^t, \quad
       \tilde \chi = \chi - \xi^t, \quad
       \tilde v = v - {k \over a} \xi^t, \quad
       \tilde v_{(i)} = v_{(i)} - {k \over a} \xi^t,
   \nonumber \\
   & & \delta \tilde \mu = \delta \mu - \dot \mu \xi^t, \quad
       \delta \tilde \mu_{(i)} = \delta \mu_{(i)} - \dot \mu_{(i)} \xi^t, \quad
       \delta \tilde p = \delta p - \dot p \xi^t, \quad
       \delta \tilde p_{(i)} = \delta p_{(i)} - \dot p_{(i)} \xi^t,
   \nonumber \\
   & & \delta \tilde Q_{(i)} = \delta Q_{(i)} - \dot Q_{(i)} \xi^t, \quad
       \tilde J_{(i)} = J_{(i)} + Q_{(i)} \xi^t.
   \label{GT}
\eea
As the temporal gauge fixing condition we can impose one condition 
in any of these temporally gauge dependent variables:
$\alpha \equiv 0$ (synchronous gauge),
$\varphi \equiv 0$ (uniform-curvature gauge),
$\kappa \equiv 0$ (uniform-expansion gauge),
$\chi \equiv 0$ (zero-shear gauge),
$v/k \equiv 0$ (comoving gauge),
$\delta \mu \equiv 0$ (uniform-density gauge), 
$\delta p \equiv 0$, $v_{(i)}/k \equiv 0$, $\delta \mu_{(i)} \equiv 0$, 
$\delta p_{(i)} \equiv 0$, etc.
By examining these we notice that,
except for the synchronous gauge condition (which fixes $\alpha = 0$), 
each of the gauge conditions fixes the temporal gauge mode completely.
Thus, a variable in such a gauge condition uniquely corresponds
to a gauge-invariant combination which combines the variable concerned
and the variable used in the gauge condition.
As examples, one can recognize the following combinations are gauge-invariant.
\bea
   & & \delta \mu_v \equiv \delta \mu - {a \over k} \dot \mu v, \quad
       \varphi_v \equiv \varphi - {aH \over k} v, \quad
       \varphi_\chi \equiv \varphi - H \chi \equiv - H \chi_\varphi, \quad
       v_\chi \equiv v - {k \over a} \chi, 
   \label{GI}
\eea
etc.
The original study of cosmological perturbation by Lifshitz 
\cite{Lifshitz-1946} was made in the synchronous gauge.
The zero-shear gauge and the comoving gauge were introduced
by Harrison \cite{Harrison-1967} and Nariai \cite{Nariai-1969}, respectively.
The combination $\varphi_v$ was first introduced by Lukash \cite{Lukash-1980}
and Bardeen \cite{Bardeen-1980}.
The gauge-invariant combinations $\varphi_\chi$ and $v_\chi$ are
equivalent to $\varphi$ and $v$ in the zero-shear gauge which takes
$\chi \equiv 0$ as the gauge condition, and
$\delta \mu_v$ is the same as $\delta \mu$ in the comoving gauge, etc.
In this way, we can systematically construct various gauge-invariant
combinations for a given variable.
Since there exist many gauge-invariant combinations even for a given
variable, say for $\delta \mu$ we could have 
$\delta \mu_\varphi$, $\delta \mu_v$, $\delta \mu_{v_{(i)}}$, etc., 
using our notation we can easily recognize which gauge-invariant 
combination we are using.
Generally, we do not know the suitable gauge condition {\it a priori}.
The proposal made in \cite{Bardeen-1988,Hwang-1991-PRW,Hwang-Noh-2001-CMBR} 
is that we write the set of equation without fixing the (temporal) gauge 
condition and arrange the equations so that we can implement easily various 
fundamental gauge conditions:
eqs. (\ref{G1}-\ref{G7}) are arranged accordingly.
We term it a {\it gauge-ready} approach.

%%%%%%%%%%%%%%%%%%%%%%%%%%%%%%%%%%%%%%%%%%%%%%%%%%%%%%%%%%%%%%%%%
\section{Fluid system}
                                                \label{sec:Adiabatic}

%%%%%%%%%%%%%%%%%%%%%%%%%%%%%%%%%%%%%%%%%%%%%%%%%%%%%%%%%%%%%%%%%
\subsection{Newtonian analog}
                                                 \label{sec:Newtonian}

\begin{figure}[ht]
   \centering
   \leavevmode
   \epsfysize=7cm
   \epsfbox{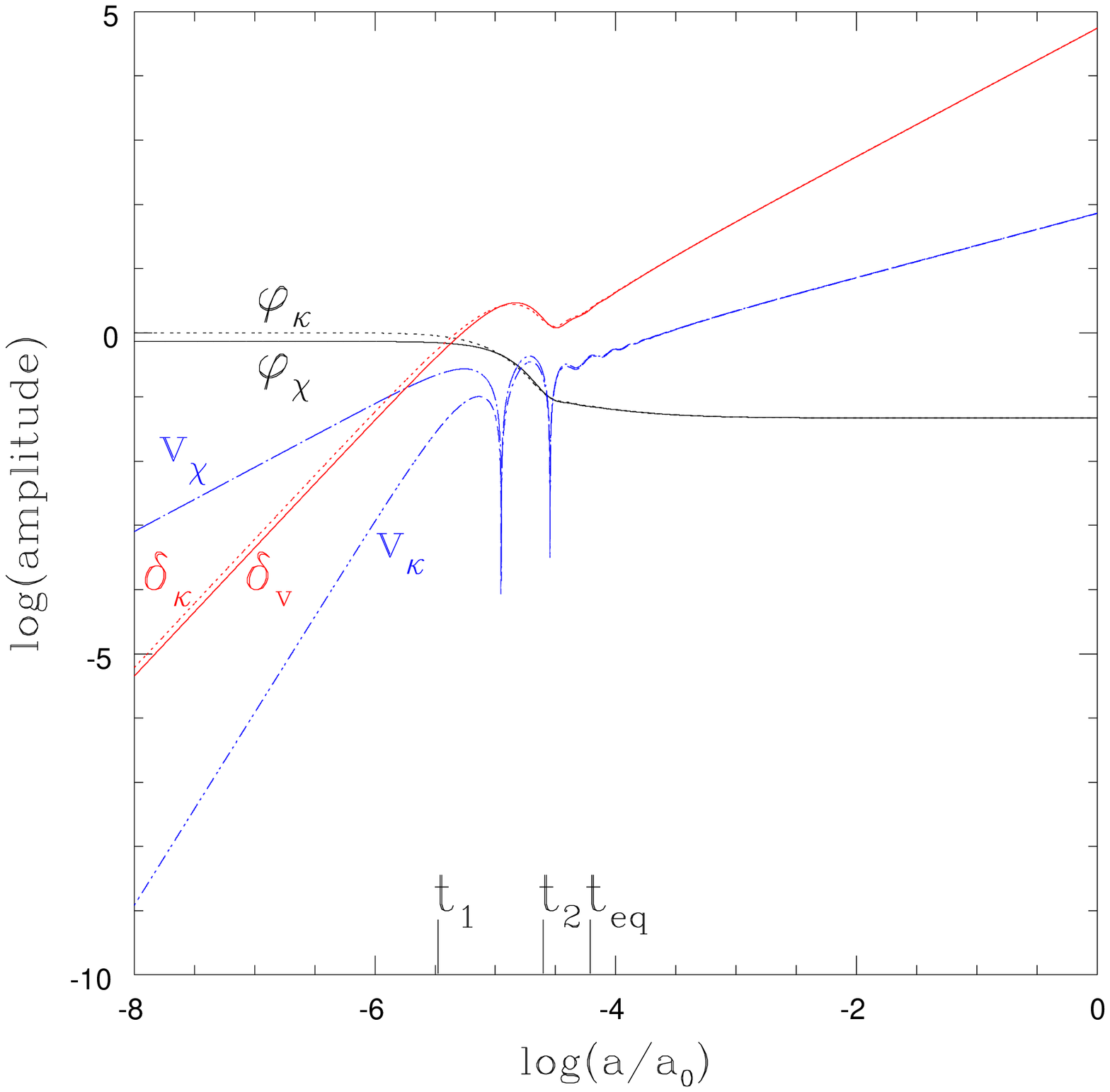}
   \epsfysize=7cm
   \epsfbox{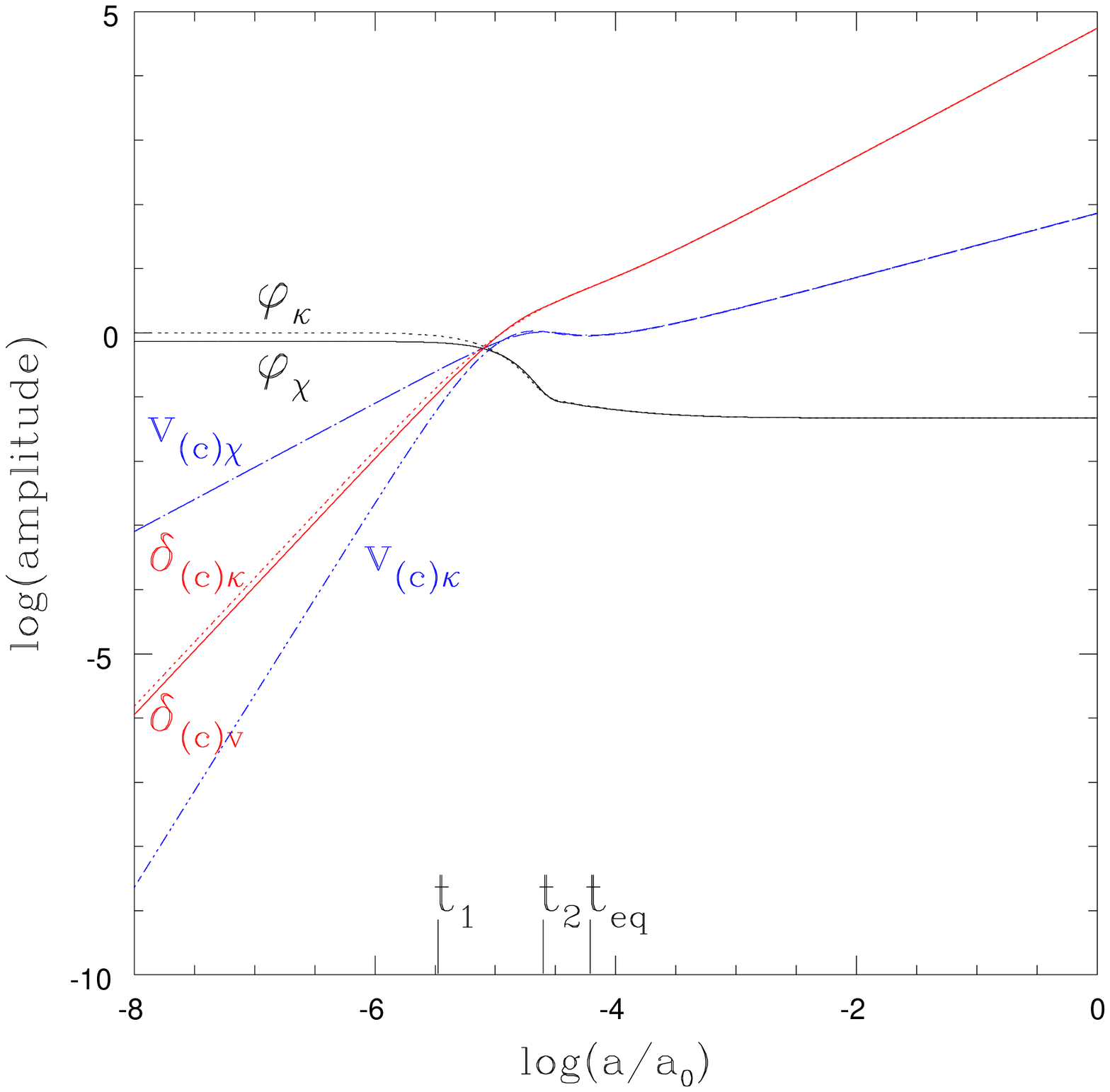}\\
   \caption[Fig-Newtonian]
   {\label{Fig-Newtonian}
In the first figure, evolutions of 
$\varphi_\chi$ (black, solid line),
${1 \over 3(1 + w)} \delta_{v}$ (red, solid line),
$v_{\chi}$ (blue, dot and long-dash line)
are compared with
$\varphi_\kappa$ (black, dotted line),
${1 \over 3(1 + w)} \delta_{\kappa}$ (red, dotted line),
$v_{\kappa}$ (blue, dot and short-dash line).
In the second figure, evolutions of 
$\varphi_\chi$ (black, solid line),
${1 \over 3} \delta_{(c)v}$ (red, solid line),
$v_{(c)\chi}$ (blue, dot and long-dash line)
are compared with
$\varphi_\kappa$ (black, dotted line),
${1 \over 3} \delta_{(c)\kappa}$ (red, dotted line),
$v_{(c)\kappa}$ (blue, dot and short-dash line).
The vertical scale indicates the relative amplitude; we normalized
$\varphi_\kappa = 1$ in the early radiation era.
We took $K = 0 = \Lambda$, $\Omega_{b0} = 0.06$, $H = 65 km/sec Mpc$,
and ignored the direct coupling between the baryon and the photon before
recombination.
Radiation-matter equality ($t_{eq}$) occurs around 
$\log(a_{eq}/a_0) \simeq -4.2$.
We took the adiabatic initial condition in the radiation dominated era.
As the scale we considered $\lambda_0 \equiv 2 \pi a_0/k = 4 \pi Mpc$.
We indicate the horizon-crossing epochs of the given scale 
as $t_1$ where $k/aH = 1$, and 
$t_2$ where $\lambda/\lambda_H = 2 \pi aH/k = 1$.
    }
\end{figure}

The following equations most closely resemble Newtonian hydrodynamic
equations in the context of linear perturbation of the Friedmann
world model:
from eqs. (\ref{G2},\ref{G3}), 
eqs. (\ref{G8},\ref{G9},\ref{G3}), 
eqs. (\ref{G4},\ref{G9}), 
eqs. (\ref{G1},\ref{G3},\ref{G4}), 
and eq (\ref{G4}), respectively we have
\bea
   & & {k^2 - 3 K \over a^2} \varphi_\chi = 4 \pi G \delta \mu_v,
   \label{Poisson-eq} \\
   & & \delta \dot \mu_v + 3 H \delta \mu_v
       = - {k^2 - 3 K \over k^2} \left[ {k \over a} (\mu + p) v_\chi
       + 2 H \pi^{(s)} \right],
   \label{delta_CG-eq} \\
   & & \dot v_\chi + H v_\chi = {k \over a} \left( - \varphi_\chi
       + {\delta p_v \over \mu + p} - 8 \pi G {a^2 \over k^2} \pi^{(s)}
       - {2 \over 3} {k^2 - 3 K \over k^2} {\pi^{(s)} \over \mu + p} \right),
   \label{v_ZSG-eq} \\
   & & \dot \varphi_\chi + H \varphi_\chi
       = - 4 \pi G {a \over k} \left( \mu + p \right) v_\chi
       - 8 \pi G H {a^2 \over k^2} \pi^{(s)},
   \label{varphi_ZSG-eq} \\
   & &  \alpha_\chi = - \varphi_\chi - 8 \pi G {a^2 \over k^2} \pi^{(s)}.
   \label{alpha-varphi}
\eea
Equations (\ref{Poisson-eq}-\ref{v_ZSG-eq}) can be
compared with the Poisson, the continuity, and the Euler (Navier-Stokes)
equations in Newtonian theory, respectively.
By setting $p = 0 = \pi^{(s)}$ and ignoring $K = 0$ we recover
the well known equations in the Newtonian perturbation theory
with $- \varphi_\chi$, $\delta \mu_v$ and $v_\chi$
corresponding to the Newtonian potential, density and velocity 
perturbations, respectively.
Notice the appearence of pressure terms in rather unexpected places
in eqs. (\ref{delta_CG-eq},\ref{v_ZSG-eq}) 
and none appearing in eq. (\ref{Poisson-eq}).
Although the zero-shear gauge and the comoving gauge were first
studied by Harrison and Nariai, respectively \cite{Harrison-1967,Nariai-1969},
these equations using mixed gauge-invariant combinations
were first presented by Bardeen in eqs. (4.3-4.8) of
\cite{Bardeen-1980}.
Newtonian correspondence of these fully general relativistic 
equations and variables was investigated in \cite{Hwang-Noh-1999-Newtonian}.
In Fig. \ref{Fig-Newtonian} we present the behaviors of 
$\varphi_\chi$, $v_\chi$ ($v_{(c)\chi}$) and $\delta_v$ ($\delta_{(c)v}$)
in the case of three component system with the photon, baryon and 
the cold dark matter ($c$).
We also compared the behavior with $\varphi$, $v$ ($v_{(c)}$) and $\delta$ 
($\delta_{(c)}$) in the uniform-expansion gauge.

In \cite{Hwang-Noh-1999-Newtonian} we noted that 
in the sub-horizon scale in matter dominated era $\varphi$, $\delta$, and $v$
in the uniform-expansion gauge all behave like $\varphi_\chi$,
$\delta_v$, and $v_\chi$ which most closely resemble the
corresponding Newtonian behavior, see also \S 84 of \cite{Peebles-1980}.
We compared the behaviors in the uniform-expansion gauge
with the variables with Newtonian behavior in Fig. \ref{Fig-Newtonian}.

%%%%%%%%%%%%%%%%%%%%%%%%%%%%%%%%%%%
%%%%%%%%%%%%%%%%%%%%%%%%%%%%%%%%%%%

{}From eqs. (\ref{G5},\ref{G8},\ref{G9}) we can derive 
\bea
   & & {1 + w \over a^2 H} \left[ {H^2 \over a (\mu + p)}
       \left( {a^3 \mu \over H} \delta_{v} \right)^\cdot \right]^\cdot
       + c_s^2 {k^2 \over a^2} \delta_{v}
       = - {k^2 - 3K \over a^2} 
       \left[ {e \over \mu} + 2 {1 + w \over H} \Big( {a^2 H^2 \over k^2} 
       {\pi^{(s)} \over \mu + p} \Big)^\cdot 
       - {2 \over 3} {\pi^{(s)} \over \mu} \right].
   \label{delta_v-eq}
\eea
The multi-component version of this equation will be presented later,
see eq. (\ref{delta_v-multi-eq}).
This equation is fully relativistic but most closely resembes the
well known Newtonian density perturbation equation; only for 
pressureless fluid our comoving gauge coincides with the synchronous gauge.
$\delta_v$ is directly related to the density gradient variable
introduced in the covariant approach \cite{Ellis-Bruni-1989}; 
see \S IV and the note added in \cite{Hwang-Vishniac-1990}.
By using eq. (\ref{Poisson-eq}), the equation for $\varphi_\chi$ 
can be derived simply from eq. (\ref{delta_v-eq}),
see eq. (\ref{varphi_chi-eq}).

In the multi-component situation, although
eqs. (\ref{delta_CG-eq},\ref{v_ZSG-eq}) are still valid for the
total fluid quantities, we can also derive the equations for
individual component.
{}From eqs. (\ref{G6},\ref{G7},\ref{G3}) and
eqs. (\ref{G7},\ref{G4}), respectively, we have
\bea
   & & \delta \dot \mu_{(i)v_{(i)}} + 3 H \delta \mu_{(i)v_{(i)}}
       = - {k^2 - 3 K \over k^2} \left[ {k \over a} (\mu_{(i)} + p_{(i)})
       v_{(i)\chi} + 2 H \pi^{(s)}_{(i)} \right]
   \nonumber \\
   & & \qquad
       + 12 \pi G {a \over k} ( \mu + p) (\mu_{(i)} + p_{(i)})
       \left( v_\chi - v_{(i)\chi} \right)
       + \delta Q_{(i)v_{(i)}} + Q_{(i)} \alpha_{v_{(i)}} 
       - 3 H J_{(i)v_{(i)}},
   \label{delta-i_CG-eq} \\
   & & \dot v_{(i)\chi} + H v_{(i)\chi} 
       + {Q_{(i)} \over \mu_{(i)} + p_{(i)}} v_{(i)\chi}
   \nonumber \\
   & & \qquad
       = {k \over a} \left( - \varphi_\chi
       + { \delta p_{(i)v_{(i)}} \over \mu_{(i)} + p_{(i)} }
       - 8 \pi G {a^2 \over k^2} \pi^{(s)}
       - {2 \over 3} {k^2 - 3 K \over k^2} 
       { \pi^{(s)}_{(i)} \over \mu_{(i)} + p_{(i)} } 
       - {J_{(i)\chi} \over \mu_{(i)} + p_{(i)} } \right).
   \label{v-i_ZSG-eq} 
\eea
By adding eqs. (\ref{delta-i_CG-eq},\ref{v-i_ZSG-eq}) over the 
components properly we have eqs. (\ref{delta_CG-eq},\ref{v_ZSG-eq}).
Later we will notice that the above equations are not only valid for
multiple fluids but also for the system with additional multiple fields.

%%%%%%%%%%%%%%%%%%%%%%%%%%%%%%%%%%%%%%%%%%%%%%%%%%%%%%%%%%%%%%%%%
\subsection{Curvature perturbations}
                                        \label{sec:Curvature-pert}

We introduce \cite{Field-Shepley-1968,Chibisov-Mukhanov-1982,%
Hwang-Vishniac-1990}
\bea
   & & \Phi \equiv \varphi_v - {K /a^2 \over 4 \pi G ( \mu + p) } \varphi_\chi,
   \label{Phi-def}
\eea
which becomes $\varphi_v$, the Lukash variable, for vanishing $K$.
Using eqs. (\ref{GI},\ref{varphi_ZSG-eq}) we can show
\bea
   & & \Phi = {H^2 \over 4 \pi G ( \mu + p) a}
       \left( {a \over H} \varphi_\chi \right)^\cdot
       + {2 H^2 \over \mu + p} {a^2 \over k^2} \pi^{(s)}.
   \label{Phi-2}
\eea
On the other hand, from eqs. (\ref{delta_v-eq},\ref{Poisson-eq},\ref{Phi-2})
or eqs. (\ref{Poisson-eq},\ref{varphi_ZSG-eq},\ref{v_ZSG-eq}) we have
\bea
   & & \dot \Phi = - {H c_s^2 \over 4 \pi G ( \mu + p) } {k^2 \over a^2}
       \varphi_\chi
       - {H \over \mu + p} \left( e - {2 \over 3} \pi^{(s)} \right).
   \label{dot-Phi-eq}
\eea
Thus, in a pressureless case, if we could ignore $e$ and $\pi^{(s)}$,
$\Phi ({\bf x}, t) = C ({\bf x})$
is an exact solution valid in general scale \cite{Field-Shepley-1968}.
Combining eqs. (\ref{Phi-2},\ref{dot-Phi-eq}) we have closed form equations
for $\Phi$ and $\varphi_\chi$:
\bea
   & & {H^2 c_s^2 \over ( \mu + p) a^3}
       \left[ {( \mu + p) a^3 \over H^2 c_s^2} \dot \Phi \right]^\cdot
       + c_s^2 {k^2 \over a^2} \Phi
       = - {H^2 c_s^2 \over (\mu + p) a^3} \left[ {a^3 \over c_s^2 H}
       \left( e - {2 \over 3} \pi^{(s)} \right) \right]^\cdot
       + 2 H^2 {c_s^2 \over \mu + p} \pi^{(s)},
   \label{Phi-eq} \\
   & & {\mu + p \over H} \left[ {H^2 \over a (\mu + p)} \left( {a \over H}
       \varphi_\chi \right)^\cdot \right]^\cdot
       + c_s^2 {k^2 \over a^2} \varphi_\chi 
       = - 4 \pi G \left( e - {2 \over 3} \pi^{(s)} \right)
       - {4 \pi G ( \mu + p) \over H} \left( {2 H^2 \over \mu + p}
       {a^2 \over k^2} \pi^{(s)} \right)^\cdot.
   \label{varphi_chi-eq}
\eea
Equation (\ref{Phi-eq}) is valid for $c_s^2 \neq 0$; for $c_s^2 = 0$ we have 
eq. (\ref{dot-Phi-eq}) instead.

In the case of multiple ideal fluids without direct interactions
among them, later we will see that the differences in the sound
velocities cause nonvanishing $e$ term which mixes the
curvature perturbation $\Phi$ with the isocurvature modes, 
see eq. (\ref{e-multi}). 
In \S \ref{sec:MSFs} we will see that a minimally coupled scalar field
has a nonvanishing entropic perturbation, see eq. (\ref{e_i-MSFs}).
Thus, we can derive a similar equation for $\Phi$ as in 
eq. (\ref{Phi-eq-MSFs}) which is the form valid even in the 
multi-component case.

Although introduced in a different way, a variable proportional to
$\Phi$ was first introduced by Field and Shepley in 1968
\cite{Field-Shepley-1968}, see also \cite{Field-1975}.
In the single component ideal fluid case, using $v \equiv z \Phi$ with
$z \equiv a \sqrt{\mu + p}/(c_s H)$
and $\prime \equiv {\partial \over \partial \eta}$, 
eq. (\ref{Phi-eq}) can be written as
\bea
   & & v^{\prime\prime} + \left( c_s^2 k^2 - z^{\prime\prime}/z \right) v = 0,
   \label{v-eq}
\eea
which is a well known equation in the cosmological perturbation,
\cite{Lukash-1980,Chibisov-Mukhanov-1982,MFB-1992,Noh-Hwang-2001-Unified}.
This equation can be found in eq. (45) of \cite{Field-Shepley-1968}
by Field and Shepley.
The variable $H$ in eq. (43) of \cite{Field-Shepley-1968}
and $\phi$ in eq. (3.3) of \cite{Chibisov-Mukhanov-1982} 
are proportional to $v$, and $\phi$ in eq. (122) of 
\cite{Hwang-Vishniac-1990} is the same as 
$-\Phi$\footnote{
       Ignorant of this rich history, and even forgetting
       our own contribution unfortunately, we have rediscovered
       this important variable $\Phi$ in \cite{Hwang-1999-Hydro}.
       }.

The $\zeta$ variable introduced in \cite{BST-1983,Bardeen-1988} is the same
as $\varphi_\delta \equiv \varphi + {\delta \mu \over 3 (\mu + p)}$.
{}From eqs. (\ref{G1},\ref{G8}) we can derive
\bea
   & & \dot \varphi_\delta = - {k \over 3 a} v_{\chi}
       - {H e \over \mu + p}.
\eea
Using eqs. (\ref{G4},\ref{G9}) we have
\bea
   & & \ddot \varphi_\delta + ( 2 - 3 c_s^2 ) H \dot \varphi_\delta
       + c_s^2 {k^2 \over a^2} \varphi_\delta
       = \left( {1 \over 3} + c_s^2 \right) {k^2 \over a^2} \varphi_\chi
   \nonumber \\
   & & \qquad
       - \left( {H e \over \mu + p} \right)^\cdot
       - \left[ ( 2 - 3 c_s^2 ) H^2 + {k^2 \over 3 a^2} \right]
       {e \over \mu + p}
       + {8 \pi G \over 3} \pi^{(s)}
       + {2 \over 9} {k^2 - 3 K \over a^2} {\pi^{(s)} \over \mu + p}.
   \label{varphi_delta-eq}
\eea
We can also derive the closed form equation for $\varphi_\delta$.
% $\ddot \varphi_\delta$ equation in a closed form?
{}From the definition of $\varphi_\delta$, evaluating it in the
comoving gauge, and using eqs. (\ref{Poisson-eq},\ref{Phi-def}) we have
\bea
   & & \varphi_\delta (k, t) = \Phi + {k^2 \over a^2}
       {1 \over 12 \pi G ( \mu + p)} \varphi_\chi,
   \label{varphi_delta-sol}
\eea
where $\Phi$ and $\varphi_\chi$ are given above.
{}For $\varphi_\kappa$ from the definition of $\varphi_\kappa$
and using eq. (\ref{G2}) and eq. (\ref{G3}) we have
\bea
   & & \varphi_\kappa = \left( 1 + {k^2 - 3 K \over 12 \pi G (\mu + p) a^2}
       \right)^{-1} \varphi_\delta
       = \varphi_\chi - \left( 1 + {k^2 - 3 K \over 12 \pi G (\mu + p) a^2}
       \right)^{-1} {aH \over k} v_\chi.
   \label{varphi_kappa}
\eea
Using the first expression in eq. (\ref{varphi_delta-eq}) 
we can guess the equation for $\varphi_\kappa$.
All the equations above are generally {\it valid} in the multi-component 
situation including fluids and fields.

In the single component ideal fluid 
(later we will see that it also applies for multiple ideal fluids
under an adiabatic condition, thus vanishing $e$) 
with ignorable $c_s^2 k^2$ term compared with $z^{\prime\prime}/z$
term in eq. (\ref{v-eq}), 
thus valid in the super-sound-horizon scale,
we have general solutions \cite{Hwang-Vishniac-1990,Hwang-Noh-1999-Newtonian}
\bea
   & & \Phi (k, t) = C (k) - d (k) {k^2 \over 4 \pi G} \int^t
       {c_s^2 H^2 \over a^3 (\mu + p)} dt,
   \label{Phi-sol} \\
   & & \varphi_\chi (k, t) = 4 \pi G C (k) {H \over a} \int^t
       {a (\mu + p) \over H^2} dt + d (k) {H \over a},
   \label{varphi_chi-sol}
\eea
where $C(k)$ and $d(k)$ are the coefficients of the growing 
and decaying solutions, respectively. 
The coefficients are matched using eqs. (\ref{Phi-2},\ref{dot-Phi-eq}).
In order to derive $\Phi$ from $\varphi_\chi$
and vice versa to proper order, we need to consider the
next leading order terms in the growing solution of $\Phi$ and
the decaying solution of $\varphi_\chi$; these follow trivially from 
eqs. (\ref{Phi-eq},\ref{varphi_chi-eq}) as
\bea
   & & \Phi = C (k) \left\{ 1 + k^2 \left[
       \int^t {a (\mu + p) \over H^2} \left(
       \int^t {c_s^2 H^2 \over a^3 (\mu + p)} dt \right) dt
       - \int^t {a (\mu + p) \over H^2} dt
       \int^t {c_s^2 H^2 \over a^3 (\mu + p)} dt \right] \right\}
   \nonumber \\
   & & \qquad
       - d (k) {k^2 \over 4 \pi G} \int^t {c_s^2 H^2 \over a^3 (\mu + p)} dt,
   \nonumber \\
   & & \varphi_\chi 
       = 4 \pi G C (k) {H \over a} \int^t {a (\mu + p) \over H^2} dt 
   \nonumber \\
   & & \qquad
       + d (k) {H \over a} \left\{ 1 + k^2 \left[
       \int^t \left( \int^t {a (\mu + p) \over H^2} dt \right) 
       {c_s^2 H^2 \over a^3 (\mu + p)} dt
       - \int^t {a (\mu + p) \over H^2} dt 
       \int^t {c_s^2 H^2 \over a^3 (\mu + p)} dt \right] \right\}.
   \label{LS-sol}
\eea
Equation (\ref{Phi-sol}) includes $c_s^2 = 0$ limit; thus for a
pressureless fluid we simply have $\Phi = C$, \cite{Field-Shepley-1968}.
Solution for $\varphi_\delta$ follows from eq. (\ref{varphi_delta-sol}).
Solution for $\varphi_\kappa$ follows using the first expression
in eq. (\ref{varphi_kappa}).
We emphasize that these solutions are valid in the super-sound-horizon
(thus virtually in all scales in the matter dominated era).

\begin{figure}[ht]
   \centering
   \leavevmode
   \epsfysize=7cm
   \epsfbox{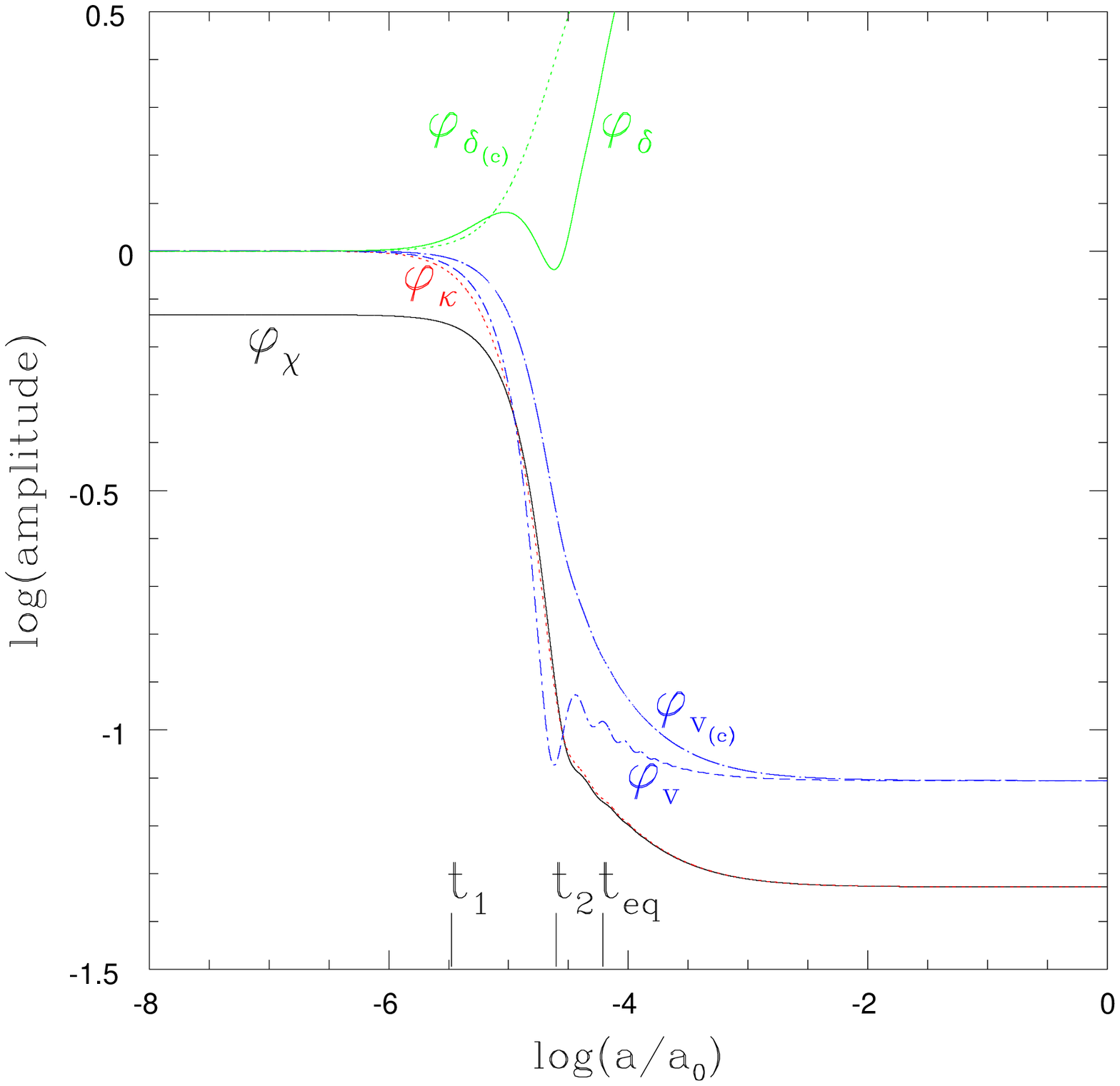}
   \epsfysize=7cm
   \epsfbox{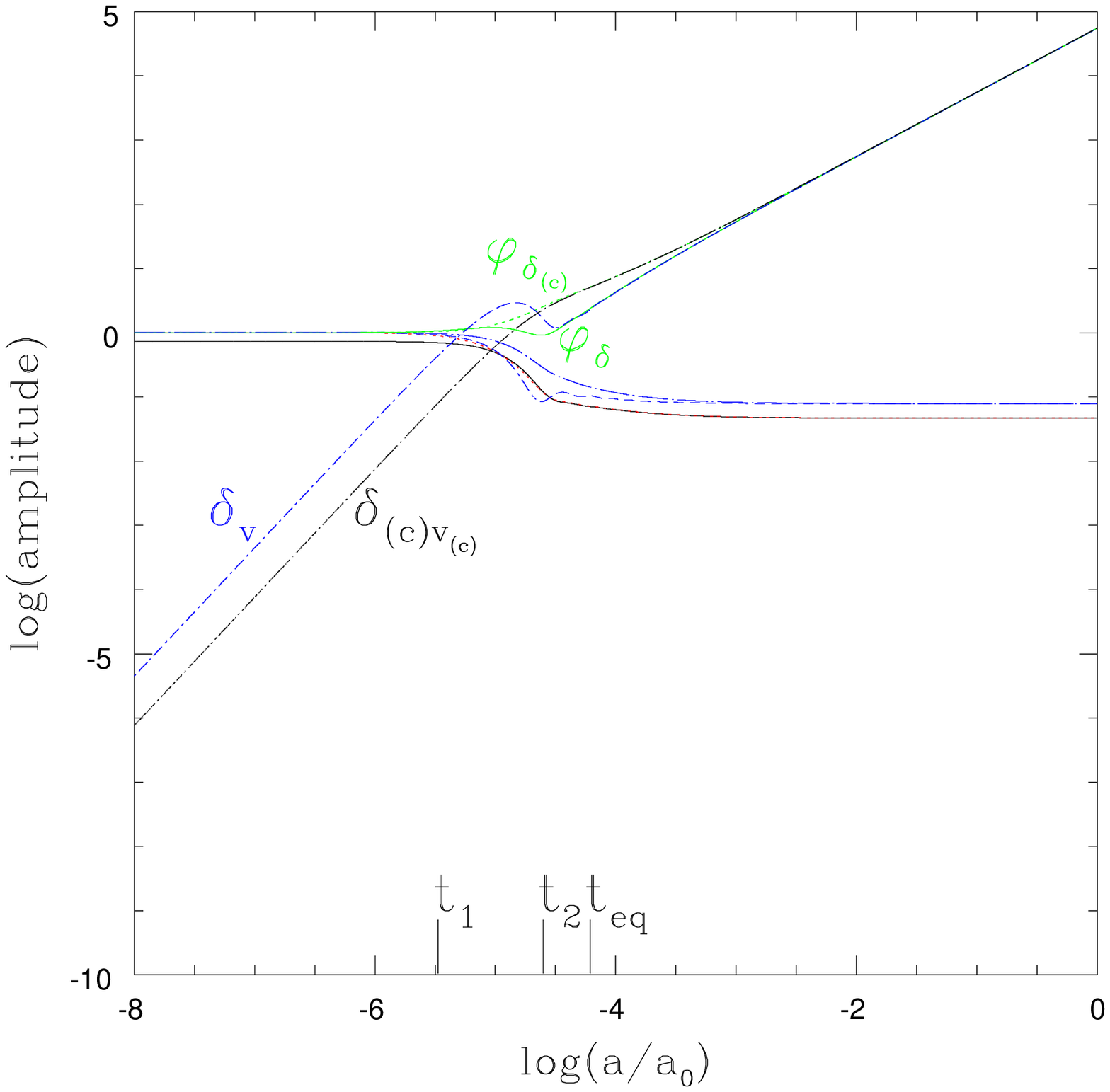}\\
   \caption[Fig-varphi]
   {\label{Fig-varphi}
The first figure shows evolutions of 
$\varphi_\chi$ (black, solid line),
$\varphi_\kappa$ (red, dotted line),
$\varphi_v$ (blue, dot and short-dash line),
$\varphi_{v(c)}$ (blue, dot and long-dash line).
The same evolutions are reproduced in the second figure
which also shows evolutions of
$\varphi_\delta$ (red, solid line),
$\varphi_{\delta_{(c)}}$ (red, dotted line),
${1 \over 3 (1 + w)} \delta_v$ (blue, dot and short-dash line),
${1 \over 3} \delta_{(c)v_{(c)}}$ (blue, dot and long-dash line).
The conditions used are the same as in Fig. \ref{Fig-Newtonian}.
    }
\end{figure}

\begin{figure}[ht]
   \centering
   \leavevmode
   \epsfysize=7cm
   \epsfbox{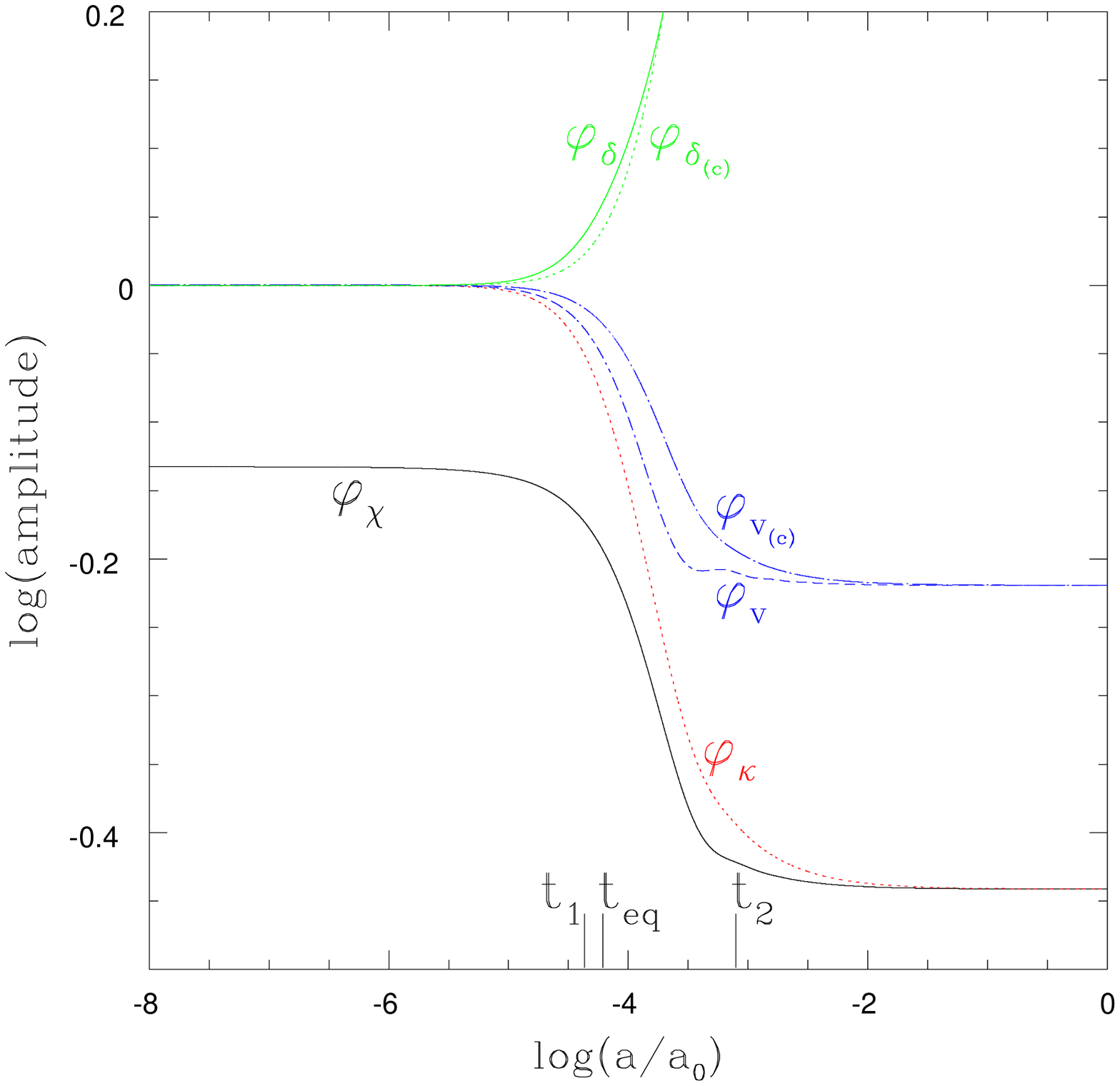}
   \epsfysize=7cm
   \epsfbox{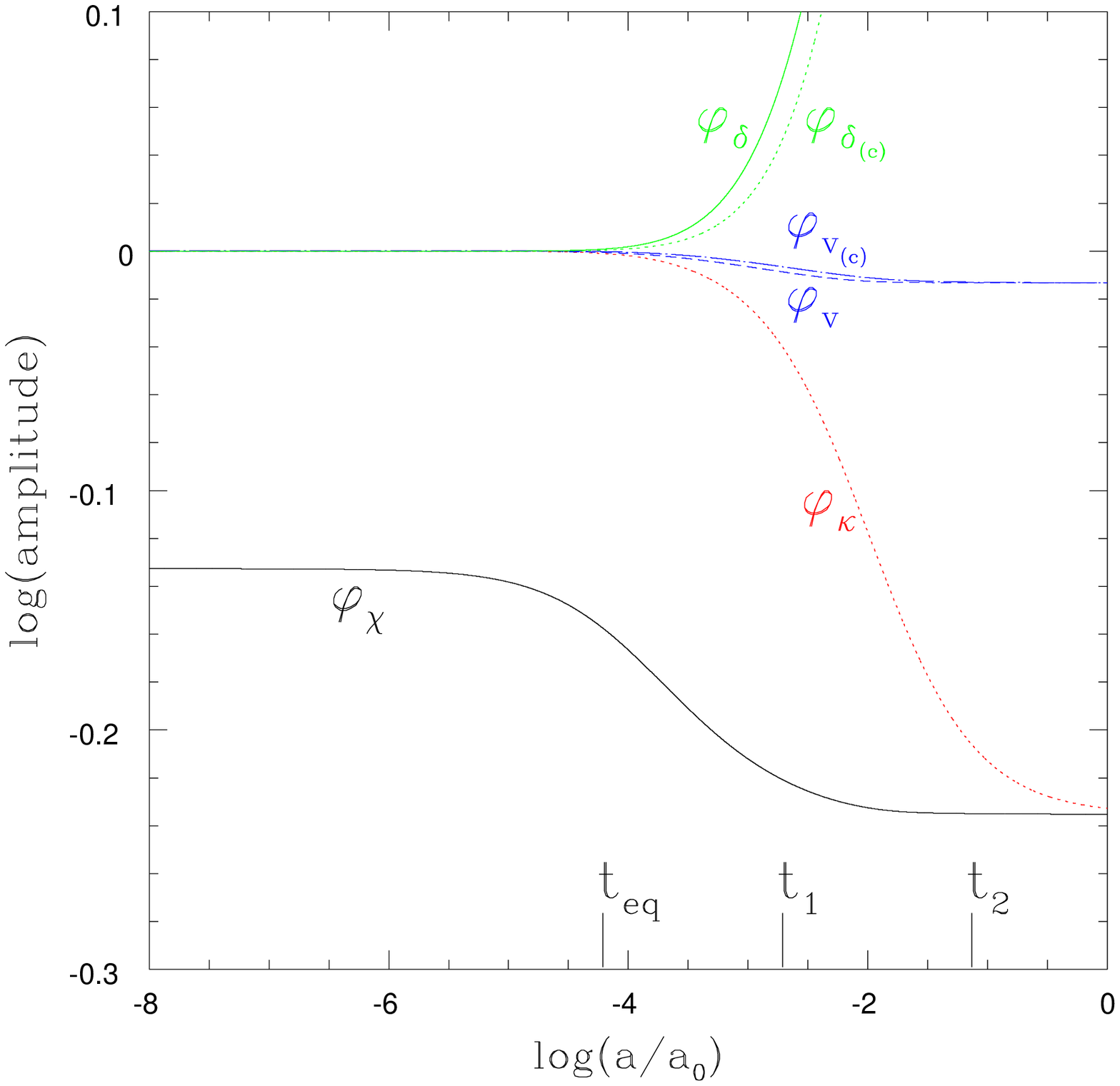}\\
   \caption[Fig-varphi-3]
   {\label{Fig-varphi-3}
The same as first figure of Fig. \ref{Fig-varphi}
for scales $\lambda_0 = 40 \pi Mpc$ and $400 \pi Mpc$, respectively.
    }
\end{figure}

Notice that the decaying solutions of $\Phi$, $\varphi_\delta$,
and $\varphi_\kappa$ (ignoring $K$ in this case) are {\it higher order} 
in the large-scale expansion compared with the one of $\varphi_\chi$.
In the super-horizon scale we have
\bea
   & & \varphi_v = \varphi_\delta = \varphi_\kappa = C,
\eea
thus time independent,
whereas $\varphi_\chi$ evolves in time according to eq. (\ref{varphi_chi-sol}).
As mentioned in \S \ref{sec:Newtonian}, despite its complex
behavior compared with the curvature variables in other gauges,
$\varphi_\chi$ most closely resembles the behavior of the
perturbed Newtonian gravitational potential \cite{Hwang-Noh-1999-Newtonian}.
In the $K = 0 = \Lambda$ background dominated by an ideal fluid
with constant $w$, in the super-sound horizon, the growing mode
behaves as
\bea
   & & \varphi_\chi = {3 + 3 w \over 5 + 3 w} C,
\eea
thus $\varphi_\chi = {2 \over 3} \varphi_v$ and
${3 \over 5} \varphi_v$ in the radiation and the matter dominated
eras, respectively.
We show the evolutions in Figs. \ref{Fig-varphi} and \ref{Fig-varphi-3}.
The amplitudes of $\varphi_v$, $\varphi_\kappa$ and $\varphi_\chi$
drop as the scale comes inside the sound-horizon;
compare Figs. \ref{Fig-varphi} and \ref{Fig-varphi-3}.

%%%%%%%%%%%%%%%%%%%%%%%%%%%%%%%%%%%
%%%%%%%%%%%%%%%%%%%%%%%%%%%%%%%%%%%
%%%%%%%%%%%%%%%%%%%%%%%%%%%%%%%%%%%
%%%%%%%%%%%%%%%%%%%%%%%%%%%%%%%%%%%

In the sub-horizon scale in matter dominated era we have the 
following behaviors:

\noindent
(i) $\varphi_v$ remains conserved.

\noindent
(ii) $\varphi_\delta$ approaches ${1 \over 3} \delta_v$.
This is because the second term in eq. (\ref{varphi_delta-sol})
dominates the first, thus
$\varphi_\delta (\equiv \varphi_v + {1 \over 3(1 +w)} \delta_v) 
\simeq {1 \over 3} \delta_v$ (in the sub-horizon $K$ term is not important);
it also follows from eq. (\ref{varphi_delta-eq}).

\noindent
(iii) $\varphi_\kappa$ approaches $\varphi_\chi$.
This follows from the second expression in eq. (\ref{varphi_kappa})
and using eq. (\ref{varphi_ZSG-eq}).

\noindent
We compared these behaviors in Fig. \ref{Fig-varphi}.
Oscillatory behaviors of $\varphi_v$ and others occur as the
scale comes inside horizon during the radiation dominated era 
where the sound-horizon scale is comparable to the (visual) horizon scale.
Since the above analytic solutions are valid for the 
super-sound-horizon we do not anticipate the above behaviors 
in (ii) and (iii) to be valid inside horizon in the radiation dominated era.
The numerical results confirm (ii) and (iii) even in such a situation: 
Case (iii) is more understandable by examining
the second expression in eq. (\ref{varphi_kappa})
and eq. (\ref{varphi_ZSG-eq}) without using
the large-scale solution.

Although many of the results in this section were presented in our
own previous work in \cite{Hwang-Noh-1999-Newtonian,Hwang-1999-Hydro},
which were mainly concerned with the single component fluid case,
we emphasize that by keeping the most general form of fluid quantities
the equations are valid even in the context of the multiple fluids
and fields as well.
{}Figures \ref{Fig-varphi} and \ref{Fig-varphi-3} show how the
conserved quantities in the single component case are affected by the
multi-component nature of the more realistic situations.

%%%%%%%%%%%%%%%%%%%%%%%%%%%%%%%%%%%%%%%%%%%%%%%%%%%%%%%%%%%%%%%%%
\subsection{Isocurvature perturbations}

We emphasize that eqs. (\ref{Poisson-eq}-\ref{varphi_kappa}) 
are generally valid for the multi-component situation.
In the multi-component case $e$ has additional contributions
besides the entropic perturbation from the individual component.
{}Following a pioneering study by Kodama and Sasaki, 
using eq. (\ref{e-def}) we decompose \cite{Kodama-Sasaki-1984}
\bea
   & & e = e_{\rm rel} + e_{\rm int}, \quad
       e_{\rm int} \equiv \sum_k e_{(k)},
   \nonumber \\
   & & e_{\rm rel} \equiv \sum_k ( c_{(k)}^2 - c_s^2 ) \delta \mu_{(k)}
       = {1 \over 2} \sum_{k,l} { ( \mu_{(k)} + p_{(k)} )
       ( \mu_{(l)} + p_{(l)} ) \over \mu + p } ( c_{(k)}^2 - c_{(l)}^2 )
       S_{(kl)}
       + \sum_k { \mu_{(k)} + p_{(k)} \over \mu + p } q_{(k)}
       c_{(k)}^2 \delta \mu,
   \label{e-multi}
\eea
where we introduced the following combinations
\bea
   & & S_{(ij)} \equiv {\delta \mu_{(i)} \over \mu_{(i)} + p_{(i)}}
       - {\delta \mu_{(j)} \over \mu_{(j)} + p_{(j)}}.
   \label{S-def}
\eea
$S_{(ij)}$ is gauge-invariant {\it for} $q_{(i)} = 0$, see eq. (\ref{GT}).
Our definition differs from a gauge-invariant definition 
based on the comoving gauge ($v/k = 0$) originally introduced in
\cite{Kodama-Sasaki-1984}, coinciding only for $q_{(i)} = 0$;
for another gauge-invariant definition see \cite{Malik-2001-Thesis}.

Thus, we notice that the RHSs of eqs. (\ref{Phi-eq},\ref{varphi_chi-eq}) 
vanish for the following situations:

\noindent
(i) single ideal fluid ($e = 0 = \pi^{(s)}$),

\noindent
(ii) multiple pressureless ideal fluids, thus $c_{(i)}^2 = 0$,

\noindent
(iii) multiple ideal fluids ($e_{(i)} = 0 = \pi_{(i)}$)
     with the $c_{(i)}^2 = c_{(j)}^2$, implying $c^2_{(i)} = c_s^2$,
     thus $e_{\rm rel} = 0$.

\noindent
(iv) multiple ideal fluids with adiabatic perturbation,
     i.e., $S_{(ij)} = 0$ for $q_{(i)} = 0$;
     for $q_{(i)} \neq 0$ it is not meaningful to set $S_{(ij)} = 0$
     without imposing the gauge condition.

In these situations the large-scale asymptotic solutions in 
eqs. (\ref{Phi-sol},\ref{varphi_chi-sol})
{\it remain valid}, thus $\Phi$ is generally conserved in the 
super-sound-horizon 
scale considering general $K$, $\Lambda$ and time-varying $p (\mu)$. 
The case (iv) shows that the strictly {\it adiabatic mode} ($S_{(ij)} = 0$)
{\it is conserved in the large-scale limit for general number of fluid 
components.}

In terms of $S_{(ij)}$ and a gauge-invariant combination
$v_{(ij)} (\equiv v_{(i)} - v_{(j)})$,
from eqs. (\ref{G6-1},\ref{G7}) we can derive \cite{Hwang-1991-PRW}
\bea
   & & \dot S_{(ij)} + {3 \over 2} H \Bigg\{ \left[ q_{(i)} ( 1 + c_{(i)}^2 )
       + q_{(j)} ( 1 + c_{(j)}^2 ) \right] S_{(ij)}
   \nonumber \\
   & & \qquad
       + \left[ q_{(i)} ( 1 + c_{(i)}^2 ) - q_{(j)} ( 1 + c_{(j)}^2 ) \right]
       \sum_k { \mu_{(k)} + p_{(k)} \over \mu + p }
       ( S_{(ik)} + S_{(jk)} ) \Bigg\}
   \nonumber \\
   & & \qquad
       = - {k \over a} v_{(ij)}
       - 3H \left[ q_{(i)} ( 1 + c_{(i)}^2 ) - q_{(j)} ( 1 + c_{(j)}^2 ) \right]
       { \delta \mu \over \mu + p }
       + 3 H ( q_{(i)} - q_{(j)} ) \alpha - 3H e_{(ij)} + \delta Q_{(ij)},
   \label{S_ij-eq} \\
   & & \dot v_{(ij)} + H v_{(ij)}
       - {3 \over 2} H \Bigg\{ \left[ c_{(i)}^2 + c_{(j)}^2
       - q_{(i)} ( 1 + c_{(i)}^2 ) - q_{(j)} ( 1 + c_{(j)}^2 ) \right] v_{(ij)}
   \nonumber \\
   & & \qquad
       + \left[ c_{(i)}^2 - c_{(j)}^2 - q_{(i)} ( 1 + c_{(i)}^2 )
       + q_{(j)} ( 1 + c_{(j)}^2 ) \right]
       \sum_k { \mu_{(k)} + p_{(k)}
       \over \mu + p} ( v_{(ik)} + v_{(jk)} ) \Bigg\}
   \nonumber \\
   & & \qquad
       = 3H \left[ c_{(i)}^2 - c_{(j)}^2 - q_{(i)} ( 1 + c_{(i)}^2 )
       + q_{(j)} ( 1 + c_{(j)}^2 ) \right] v
       + {k \over a} \Bigg[
       ( c_{(i)}^2 - c_{(j)}^2 ) { \delta \mu \over \mu + p } 
       + {1 \over 2} ( c_{(i)}^2 + c_{(j)}^2 ) S_{(ij)}
   \nonumber \\
   & & \qquad
       + {1 \over 2} ( c_{(i)}^2 - c_{(j)}^2 ) \sum_k {\mu_{(k)} + p_{(k)}
       \over \mu + p} ( S_{(ik)} + S_{(jk)} )
       + e_{(ij)} - {2 \over 3} {k^2 - 3K \over k^2} \pi^{(s)}_{(ij)}
       - J_{(ij)} \Bigg],
   \label{v_ij-eq}
\eea
where $e_{(ij)}$, $\pi^{(s)}_{(ij)}$, $\delta Q_{(ij)}$, and $J_{(ij)}$
are defined similarly as $S_{(ij)}$:
\bea
   & & e_{(ij)} \equiv {e_{(i)} \over \mu_{(i)} + p_{(i)}}
       - {e_{(j)} \over \mu_{(j)} + p_{(j)}}, \quad
       \pi^{(s)}_{(ij)} \equiv {\pi^{(s)}_{(i)} \over \mu_{(i)} + p_{(i)}}
       - {\pi^{(s)}_{(j)} \over \mu_{(j)} + p_{(j)}}, 
   \nonumber \\
   & & \delta Q_{(ij)} \equiv {\delta Q_{(i)} \over \mu_{(i)} + p_{(i)}}
       - {\delta Q_{(j)} \over \mu_{(j)} + p_{(j)}}, \quad
       J_{(ij)} \equiv {J_{(i)} \over \mu_{(i)} + p_{(i)}}
       - {J_{(j)} \over \mu_{(j)} + p_{(j)}}.
   \label{e_ij-def}
\eea
Although $S_{(ij)}$, $\delta Q_{(ij)}$ and $J_{(ij)}$
are not gauge-invariant for nonvanishing $q_{(i)}$, 
eqs. (\ref{S_ij-eq},\ref{v_ij-eq}) are written in a gauge-ready form;
these equations were derived in eqs. (35,36) of \cite{Hwang-1991-PRW}.
Under the comoving gauge condition based on $v/k = 0$,
removing $\alpha_v$ using eq. (\ref{G9}), we can show that
eqs. (\ref{S_ij-eq},\ref{v_ij-eq}) reduce to 
(the corrected) eqs. (5.53,5.57) in \cite{Kodama-Sasaki-1984};
typographical errors in \cite{Kodama-Sasaki-1984}
were corrected in eqs. (A.37,A.38) of \cite{Hamazaki-Kodama-1996}.
We emphasize that, since eqs. (\ref{G6},\ref{G7}) are valid in the
context of generalized gravity theories considered in 
\cite{Hwang-Noh-2001-CMBR}, eqs. (\ref{S_ij-eq},\ref{v_ij-eq})
in the same forms remain valid in the context of the
generalized gravity theories as well.

In the case of multiple field system, we have nonvanishing
entropic perturbation of the individual component.
Later we will show that the relative entropic perturbation
$e_{(ij)}$ causes a curious situation of making eq. (\ref{v_ij-eq})
simple, and in fact making it identically satisfied, and 
eq. (\ref{S_ij-eq}) alone gives the second order system; 
see eqs. (\ref{e_ij-MSFs}-\ref{delta-phi_ij-eq}).

{}For $q_{(i)} = 0$, $S_{(ij)}$ is gauge-invariant.
In such a case 
eqs. (\ref{S_ij-eq},\ref{v_ij-eq},\ref{delta_v-eq},\ref{e-multi}) 
form a closed set of equations for $S_{(ij)}$ and $\delta_v$;
$\delta \mu$ and $v$ term in eq. (\ref{v_ij-eq})
combine to give $\delta \mu_v$.
In this way, $\delta \mu_v$ and $S_{(ij)}$ may provide the curvature and
isocurvature perturbations, respectively.
Using eqs. (\ref{Poisson-eq},\ref{dot-Phi-eq}) we can express
$\delta \mu_v$ in terms $\Phi$ and $\varphi_\chi$ as
\bea
   & & \delta \mu_v = {k^2 - 3 K \over 4 \pi G a^2} \varphi_\chi
       = - {1 - 3K/k^2 \over c_s^2} 
       \left( {\mu + p \over H} \dot \Phi + e - {2 \over 3} \pi^{(s)} \right),
   \label{delta-mu_v}
\eea
where $e$ is given in eq. (\ref{e-multi}),
and eqs. (\ref{Phi-eq},\ref{varphi_chi-eq}) provide the equation for $\Phi$
and $\varphi_\chi$.
In this way, $\Phi$ (or $\varphi_\chi$) 
and $S_{(ij)}$ may provide the curvature and isocurvature perturbations. 
Our use of these nomenclature 
is somewhat misleading in the sense that, actually, 
the conditions $S_{(ij)} = 0$ make the perturbation adiabatic, 
and the condition $\Phi = 0$ (or $\varphi_\chi = 0$)
makes the perturbations isocurvature.
[In the literature, often $\varphi_\chi$ is chosen to indicate the
curvature mode.
$\varphi_\chi$ is simply related to $\delta \mu_v$ as in eqs.
(\ref{Poisson-eq}), and it is related to $\Phi$ as in eq. (\ref{Phi-2}).]
We consider $\Phi$ (or $\varphi_\chi$) and $S_{(ij)}$ characterize
the curvature (adiabatic) and isocurvature perturbations, respectively.
{}For nonvanishing $q_{(i)}$ we still have a freedom to impose a
temporal gauge-condition depending on the situation, and we can use
eqs. (\ref{G1}-\ref{G5}) to replace the metric and matter variables
in terms of the relevant curvature and isocurvature perturbation variables.
{}For $n$-component fluids we can derive $n$-set of coupled second-order
differential equations.

In the case we could {\it ignore} the mutual interaction among the
fluids for the background (thus, set $q_{(i)} = 0$)
eqs. (\ref{S_ij-eq},\ref{v_ij-eq}) give
\bea
   & & \ddot S_{(ij)} + \left[ 2 - {3 \over 2} ( c_{(i)}^2 + c_{(j)}^2) 
       \right] H \dot S_{(ij)}
       - {3 \over 2} H ( c_{(i)}^2 - c_{(j)}^2 )
       \sum_{k} {\mu_{(k)} + p_{(k)} \over \mu + p}
       \left( \dot S_{(ik)} + \dot S_{(jk)} \right)
   \nonumber \\
   & & \qquad
       + {1 \over 2} {k^2 \over a^2} \Bigg[
       ( c_{(i)}^2 + c_{(j)}^2 ) S_{(ij)}
       + ( c_{(i)}^2 - c_{(j)}^2 )
       \sum_{k} {\mu_{(k)} + p_{(k)} \over \mu + p}
       \left( S_{(ik)} + S_{(jk)} \right) \Bigg]
   \nonumber \\
   & & \qquad
       = - {k^2 \over a^2} ( c_{(i)}^2 - c_{(j)}^2 )
       {\delta \mu_v \over \mu + p}
       - \left( 3 H e_{(ij)} - \delta Q_{(ij)} \right)^\cdot
       - \left[ 2 - {3 \over 2} ( c_{(i)}^2 + c_{(j)}^2 ) \right] 
       H \left( 3 H e_{(ij)} - \delta Q_{(ij)} \right)
   \nonumber \\
   & & \qquad
       + {3 \over 2} H ( c_{(i)}^2 - c_{(j)}^2 )
       \sum_k {\mu_{(k)} + p_{(k)} \over \mu + p}
       \Big[ 3 H ( e_{(ik)} + e_{(jk)} )
       - \delta Q_{(ik)} - \delta Q_{(jk)} \Big]
   \nonumber \\
   & & \qquad
       - {k^2 \over a^2} \left( e_{(ij)} 
       - {2 \over 3} {k^2 - 3 K \over k^2} \pi_{(ij)}^{(s)}
       - J_{(ij)} \right),
   \label{S_ij-eq-2}
\eea
where $\delta \mu_v$ is given in eq. (\ref{delta-mu_v});
equations for $\delta \mu_v$, $\Phi$ and $\varphi_\chi$ are given 
in eqs. (\ref{delta_v-eq},\ref{Phi-eq},\ref{varphi_chi-eq}).
{\it Assuming} no mutual interaction among fluids (thus,
$\delta Q_{(i)} = 0 = J_{(i)}$)
and ideal fluids (thus, $e_{(i)} = 0 = \pi_{(i)}^{(s)}$),
the term in the RHS of eq. (\ref{S_ij-eq-2}) becomes
\bea
   & & - {k^2 \over a^2} ( c_{(i)}^2 - c_{(j)}^2 )
       {\delta \mu_v \over \mu + p},
   \label{S_ij-eq-2-RHS}
\eea
which vanishes for

\noindent
(i) pressureless fluids,

\noindent
(ii) fluids with $c_{(i)}^2 = c_{(j)}^2$,

\noindent
(iii) $\delta \mu_v$ = 0,

\noindent
(iv) most importantly, {\it by comparing} 
eqs. (\ref{S_ij-eq-2},\ref{delta-mu_v},\ref{delta_v-eq},\ref{e-multi})
or eqs. (\ref{S_ij-eq-2},\ref{delta-mu_v},\ref{Phi-eq},\ref{e-multi}), 
we notice that, 
for $K = 0$, the RHS becomes negligible in the large-scale limit.
Therefore, in the large-scale limit the curvature mode in the
RHS of eq. (\ref{S_ij-eq-2}) does not contribute to the isocurvature modes,
thus, isocurvature modes {\it decouple} from the curvature one;
this result was known by Kodama and Sasaki, see eq. (2.16) 
in \cite{Kodama-Sasaki-1986};
for another argument in a two-component situation,
see \cite{Wands-etal-2000}.
This should be compared with the scalar field system where
we expect less strong decoupling in general;
see below eq. (\ref{delta-mu_v-MSFs}) and \cite{Hwang-Noh-2000-MSFs}.

In the cases of (i) and (ii), assuming ideal fluids
($e_{(ij)} = 0 = \pi_{(ij)}^{(s)}$) and without interactions among
fluids ($\delta Q_{(ij)} = 0 = J_{(ij)}$),
eq. (\ref{S_ij-eq-2}) becomes
\bea
   & & \ddot S_{(ij)} + \left( 2 - 3 c_s^2 \right) H \dot S_{(ij)}
       + c_s^2 {k^2 \over a^2} S_{(ij)} = 0,
   \label{S_ij-eq-3}
\eea
where $c_s^2 \equiv c_{(i)}^2$.
In such a case the RHS of eq. (\ref{delta_v-eq}) vanishes,
thus we have a general large-scale solution for $\delta_v$ 
\bea
   & & \delta_v = {1 \over \Omega} \sum_{k} \Omega_{(k)} \delta_{(k)v}
       = (k^2 - 3 K) {H \over a^3 \mu} \left[ C(k) \int^t
       {a ( \mu + p) \over H^2} dt + {1 \over 4 \pi G} d (k) \right],
\eea
where the coefficients are matched using
eqs. (\ref{Poisson-eq},\ref{varphi_chi-sol}).
If $w = w_{(i)} = {\rm constant}$ we have $c_s^2 = w$, 
in the large-scale limit the general solution of eq. (\ref{S_ij-eq-3}) becomes
\bea
   & & S_{(ij)} = C_{ij} (k) + D_{ij} (k) \int^t {dt \over a^{2-3w}}.
   \label{S_ij-eq-3-sol}
\eea
We note that although equations in terms of $\delta_v$
and $S_{(ij)}$ are decoupled in this case, the equations 
directly in terms of $\delta_{(i)}$ are generally coupled
even in the same situation, see eq. (\ref{delta_v-multi-eq}) as an example;
similar situations in the scalar field system can be found
below eq. (\ref{delta-phi-i-eq-SR}).

In the two-component ideal fluid system without mutual interactions, 
using $S \equiv S_{(12)}$, eq. (\ref{S_ij-eq-2}) becomes
\bea
   & & \ddot S + H \left ( 2 - 3 c_z^2 \right) \dot S
       + c_z^2 {k^2 \over a^2} S = - {k^2 \over a^2} ( c_{(1)}^2
       - c_{(2)}^2 ) {\delta \mu_v \over \mu + p},
   \nonumber \\
   & & \qquad
       c_z^2 \equiv {c_{(2)}^2 ( \mu_{(1)} + p_{(1)} )
       + c_{(1)}^2 ( \mu_{(2)} + p_{(2)} ) \over \mu + p}.
   \label{S-eq-two}
\eea
Various asymptotic solutions in the case of radiation (ideal fluid with 
$p_{(r)} = {1 \over 3} \mu_{(r)}$)
plus dust (pressureless ideal fluid with $p_{(c)} = 0$) 
system were studied thoroughly by Kodama and Sasaki in 
\cite{Kodama-Sasaki-1986,Kodama-Sasaki-1987}.

The adiabatic and the various isocurvature initial conditions
in a realistic four component system with the cold dark matter,
the massless neutrinos and the tighly coupled baryon and photon
in the radiation dominated limit are presented in \cite{Bucher-etal-2000}.

%%%%%%%%%%%%%%%%%%%%%%%%%%%%%%%%%%%%%%%%%%%%%%%%%%%%%%%%%%%%%%%%%
\subsection{Component equations}
                                          \label{sec:component}

In this subsection we {\it assume} no interaction among fluids, 
thus $q_{(i)} = 0$ and $\delta Q_{(i)} = 0 = J_{(i)}$.
We have
\bea
   & & \varphi_{\delta_{(i)}} 
       \equiv \varphi - H {\delta \mu_{(i)} \over \dot \mu_{(i)}}
       = \varphi + {\delta \mu_{(i)} \over 3 (\mu_{(i)} + p_{(i)})},
   \label{varphi_delta_i-def}
\eea
where in the second step we used $q_{(i)} = 0$.
{}From eqs. (\ref{G1},\ref{G6}) we can derive
\bea
   & & \dot \varphi_{\delta_{(i)}} = - {k \over 3 a} v_{(i)\chi}
       - {H e_{(i)} \over \mu_{(i)} + p_{(i)}}.
\eea
Using eqs. (\ref{G4},\ref{G7}) we can derive
\bea
   & & \ddot \varphi_{\delta_{(i)}} + H ( 2 - 3 c_{(i)}^2 )
       \dot \varphi_{\delta_{(i)}}
       + c_{(i)}^2 {k^2 \over a^2} \varphi_{\delta_{(i)}}
       = \left( {1 \over 3} + c_{(i)}^2 \right) {k^2 \over a^2} \varphi_\chi
   \nonumber \\
   & & \qquad
       - \left( {H e_{(i)} \over \mu_{(i)} + p_{(i)}} \right)^\cdot
       - \left[ ( 2 - 3 c_{(i)}^2 ) H^2 + {k^2 \over 3 a^2} \right]
       {e_{(i)} \over \mu_{(i)} + p_{(i)}}
       + {8 \pi G \over 3} \pi^{(s)}
       + {2 \over 9} {k^2 - 3 K \over a^2} 
       {\pi_{(i)}^{(s)} \over \mu_{(i)} + p_{(i)}}.
   \label{varphi_delta_i-eq} 
\eea
Using $\varphi_{\delta_{(i)}} = {\delta_{(i)\varphi} \over 3 (1 + w_{(i)})}$ 
eq. (\ref{varphi_delta_i-eq}) can be directly converted to a set of
equations for $\delta_{(i)\varphi}$, i.e., equations for $\delta_{(i)}$
in the uniform-curvature gauge.
% $\ddot \varphi_{\delta_{(i)}}$ equations, closed form?
% $\ddot \varphi_{v_{(i)}}$ equations, simple and closed forms?
We plot the evolution of $\varphi_{\delta_{(c)}}$ for the
cold dark matter $(c)$ in Figs. \ref{Fig-varphi} and \ref{Fig-varphi-3}.
During the super-horizon scale $\varphi_{\delta_{(c)}}$ and
$\varphi_\delta$ are conserved like $\varphi_v$ and others.
As they come inside the horizon they are dominated by the
density parts; in the matter dominated era using 
eqs. (\ref{varphi_delta_i-eq},\ref{varphi_delta-eq}) 
we can show $\varphi_{\delta_{(c)}}$ and $\varphi_\delta$ behave
as ${1 \over 3} \delta_v \simeq {1 \over 3} \delta_{(c)v}
\simeq {1 \over 3} \delta_{(c)v_{(c)}}$.

It is also convenient to have equations for the individual component
of density perturbation in a gauge-invariant combination
with the velocity of the same component, i.e., in terms of
$\delta_{(i)v_{(i)}} \equiv \delta_{(i)} + 3 H(a/k) ( 1 + w_{(i)} ) v_{(i)}$.
By evaluating eqs. (\ref{G5},\ref{G6},\ref{G7}) in the $v_{(i)} = 0$ gauge
condition we can derive 
\bea
   & & \ddot \delta_{(i)v_{(i)}} + \left[ 2 + 3 ( c_{(i)}^2 - 2 w_{(i)} )
       \right] H \dot \delta_{(i)v_{(i)}}
       + \Big[ c_{(i)}^2 {k^2 \over a^2} 
       - 3 \dot H ( w_{(i)} + c_{(i)}^2 )
       + 3 H^2 ( 3 c_{(i)}^2 - 5 w_{(i)} ) \Big]
       \delta_{(i)v_{(i)}}
   \nonumber \\
   & & \qquad
       - 4 \pi G ( 1 + w_{(i)} ) \sum_k \mu_{(k)} 
       ( 1 + 3 c_{(k)}^2 ) \delta_{(k)v_{(i)}}
%   \nonumber \\
%   & & \qquad
%       = {1 + w_{(i)} \over a^2 H} 
%       \left[ {H^2 \over a ( \mu_{(i)} + p_{(i)} ) }
%       \left( {a^3 \mu_{(i)} \over H} \delta_{(i)v_{(i)}} \right)^\cdot
%       \right]^\cdot
%       + c_{(i)}^2 {k^2 \over a^2} \delta_{(i)v_{(i)}}
%   \nonumber \\
%   & & \qquad
%       + 4 \pi G \Big[ ( 1 + w ) \mu ( 1 + 3 c_s^2 ) \delta_{(i)v_{(i)}}
%       - ( 1 + w_{(i)} ) \sum_k \mu_{(k)}
%       ( 1 + 3 c_{(k)}^2 ) \delta_{(k) v_{(i)}} \Big]
   \nonumber \\
   & & = {12 \pi G \over \mu_{(i)}} \sum_k \Big[ 
       ( \mu_{(i)} + p_{(i)} ) e_{(k)}
       - \left( \mu_{(k)} + p_{(k)} \right) e_{(i)} \Big]
       - {k^2 - 3 K \over a^2} {e_{(i)} \over \mu_{(i)}}
   \nonumber \\
   & & \qquad
       - {k^2 - 3 K \over k^2} \Bigg[ 2 {1 + w_{(i)} \over a^2 H}
       \left( {a^2 H^2 \pi_{(i)}^{(s)} \over \mu_{(i)} + p_{(i)}} \right)^\cdot
       - {2 \over 3} {k^2 \over a^2} {\pi_{(i)}^{(s)} \over \mu_{(i)}} \Bigg],
   \label{delta_v-multi-eq}
\eea
with
\bea
   & & 4 \pi G ( 1 + w_{(i)} ) \sum_k \mu_{(k)} 
       ( 1 + 3 c_{(k)}^2 ) \delta_{(k)v_{(i)}}
       = 4 \pi G ( 1 + w_{(i)} ) \sum_k \mu_{(k)}
       ( 1 + 3 c_{(k)}^2 ) \delta_{(k)v_{(k)}}
       + {1 2 \pi G H \over 3 \dot H - k^2/a^2} \sum_k
   \nonumber \\
   & & \qquad
       \times \mu_{(k)} ( 1 + 3 c_{(k)}^2 )
       \Big[ ( 1 + w_{(k)} ) \left( \dot \delta_{(i)v_{(i)}}
       - 3 H w_{(i)} \delta_{(i)v_{(i)}} \right)
       - ( 1 + w_{(i)} ) \left( \dot \delta_{(k)v_{(k)}}
       - 3 H w_{(k)} \delta_{(k)v_{(k)}} \right) \Big],
   \label{relation-1}
\eea
where we used
\bea
   & & \left( 3 \dot H - {k^2 \over a^2} \right) {a \over k} v_{(ij)}
      =  {1 \over 1 + w_{(i)}} \left( \dot \delta_{(i)v_{(i)}}
       - 3 H w_{(i)} \delta_{(i)v_{(i)}} \right)
       - {1 \over 1 + w_{(j)}} \left( \dot \delta_{(j)v_{(j)}}
       - 3 H w_{(j)} \delta_{(j)v_{(j)}} \right),
   \label{v_ij-relation}
\eea
which follows from eq. (\ref{G6})  
using $\kappa_{v_{(i)}} = \kappa + (3 \dot H - k^2/a^2) (a/k) v_{(i)}$.
We have
\bea
   & & \delta_v = {1 \over \mu} \sum_k \mu_{(k)} \delta_{(k)v_{(k)}}, \quad
       S_{(ij)} = 
       {\delta \mu_{(i) v_{(i)}} \over \mu_{(i)} + p_{(i)}}
       - {\delta \mu_{(j) v_{(j)}} \over \mu_{(j)} + p_{(j)}}
       - 3 H {a \over k} v_{(ij)}.
   \label{delta_v-S_ij}
\eea
Equations (\ref{delta_v-multi-eq},\ref{relation-1}) 
provide a closed set of second-order
differential equations in terms of $\delta_{(i) v_{(i)}}$'s.
This includes the case with one additional scalar field,
see eq. (\ref{e_i-MSFs}) where 
$\delta \mu_{(i)v_{(i)}} = \delta \mu_{(i)\delta \phi_{(i)}}
= - \dot \phi^2_{(i)} \alpha_{\delta \phi_{(i)}}$;
for multiple scalar fields we also have nonvanishing
$\delta Q_{(i)}$ in eq. (\ref{delta-Q-MSFs}). 
In the single component situation $\delta_{(i)v_{(i)}}$ becomes
$\delta_v$ which behaves similarly as in the Newtonian situation
\cite{Hwang-Noh-1999-Newtonian}.
In the single component limit eq. (\ref{delta_v-multi-eq})
reduces to eq. (\ref{delta_v-eq}).
Using eqs. (\ref{G1}-\ref{G9}) one can also derive a closed form 
equations in terms of $\delta_{(i)v}$.

{}For pressureless ideal fluids 
($w_{(i)} = 0$ and $e_{(i)} = 0 = \pi_{(i)}^{(s)}$) 
eq. (\ref{delta_v-multi-eq}) gives
\bea
   & & \ddot \delta_{(i)} + 2 H \dot \delta_{(i)} 
       + c_{(i)}^2 {k^2 \over a^2} \delta_{(i)}
       - 4 \pi G \sum_k \mu_{(k)} \delta_{(k)} = 0,
   \label{presureless-delta-multi-eq}
\eea
under the $v_{(i)} = 0$ gauge condition; thus the complete
set of equations is made under mixed gauge conditions.
Using eqs. (\ref{G1}-\ref{G9}) we can show that
eq. (\ref{presureless-delta-multi-eq}) is valid
under the $v_{(j)} = 0$ or $v = 0$ gauge conditions as well.
Thus we can regard eq. (\ref{presureless-delta-multi-eq}) is valid
under {\it any} single comoving gauge condition of
$v_{(j)} = 0$ or $v = 0$; the $v_{(c)} = 0$ gauge is the same as
the synchronous gauge without remaining gauge mode.
In fact, one can derive eq. (\ref{presureless-delta-multi-eq}) 
in the Newtonian context.

In the case of ideal fluids with one component a pressureless fluid
($w_{(c)} = 0$), eq. (\ref{delta_v-multi-eq}) for $i = c$ becomes
\bea
   & & \ddot \delta_{(c)v_{(c)}} + 2 H \dot \delta_{(c)v_{(c)}} 
       - 4 \pi G \sum_k \mu_{(k)} (1 + 3 c_{(k)}^2) \delta_{(k)v_{(k)}} = 0.
\eea
Solutions in the case of radiation and dust were studied
in \cite{r-d-solutions}.

%%%%%%%%%%%%%%%%%%%%%%%%%%%%%%%%%%%%%%%%%%%%%%%%%%%%%%%%%%%%%%
\section{Scalar field system}
                                                \label{sec:MSFs}

Considering our previous study of the multiple scalar field system 
in \cite{Hwang-Noh-2000-MSFs} the following can be considered as 
a supplement to \cite{Hwang-Noh-2000-MSFs}.
Besides considering general $K$ and $\Lambda$ the equations are
presented in the context of fluid formulation using
fluid quantities; in fact, part of the fluid formulation was presented in
Sec. 4.4.2 of \cite{Hwang-1991-PRW}.
By presenting the scalar field system together with the real fluid system
we can compare the similarity and difference of the two systems
more easily.

A system of minimally coupled scalar fields can be re-interpreted 
as a system of fluids with special fluid quantities.
{}For an arbitrary number of minimally coupled scalar fields we have
\bea
   & & T_{ab} = \sum_{k} \left( \phi_{(k),a} \phi_{(k),b}
       - {1 \over 2} g_{ab} \phi_{(k)}^{\;\;\;\; ;c} \phi_{(k),c} \right)
       - V g_{ab},
   \label{Tab-MSFs}
\eea
where $\phi_{(i)}$ indicates $i$-th scalar field with
$i, j, k, \dots = 1, 2, \dots n$;
$V = V(\phi_{(k)}) = V (\phi_{(1)}, \phi_{(2)}, \dots , \phi_{(n)})$.
Additionally we have equations of motion for the scalar fields as
\bea
   & & \phi^{\;\;\;\; ;a}_{(i)\;\;\; a} - V_{,(i)} = 0,
   \label{EOM-MSFs}
\eea
where $V_{,(i)} \equiv \partial V / (\partial \phi_{(i)})$.

We decompose the scalar fields as
$\phi_{(i)} = \bar \phi_{(i)} + \delta \phi_{(i)}$.
To the background order, from eqs. (\ref{Tab},\ref{Tab-MSFs},\ref{EOM-MSFs})
we have
\bea
   & & \mu = {1 \over 2} \sum_{(k)} \dot \phi_{(k)}^2 + V, \quad
       p = {1 \over 2} \sum_{(k)} \dot \phi_{(k)}^2 - V, 
   \label{MSFs-BG} \\
   & & \ddot \phi_{(i)} + 3 H \dot \phi_{(i)} + V_{,(i)} = 0.
   \label{EOM-MSFs-BG}
\eea
Equations (\ref{BG1},\ref{BG2},\ref{MSFs-BG},\ref{EOM-MSFs-BG}) 
provide a complete set for the background evolution.
In order to match with eq. (\ref{BG2}) 
it is convenient to introduce fluid quantities of the individual field as
\bea
   & & \mu_{(i)} + p_{(i)} \equiv \dot \phi_{(i)}^2, \quad
       \dot \mu_{(i)} \equiv - 3 H \left( \mu_{(i)} + p_{(i)} \right).
   \label{MSFs-BG-fluid}
\eea
In this way we have $Q_{(i)} = 0$ by definition.
{}From eq. (\ref{MSFs-BG-fluid}) we have
$\dot \mu_{(i)} = ( \ddot \phi_{(i)} + V_{,(i)} ) \dot \phi_{(i)}$ and
$\dot p_{(i)} = ( \ddot \phi_{(i)} - V_{,(i)} ) \dot \phi_{(i)}$.

To the perturbed order, from eqs. (\ref{Tab},\ref{Tab-MSFs}) we have
\bea
   & & \delta \mu = \sum_{k} \left( \dot \phi_{(k)} \delta \dot \phi_{(k)}
       - \dot \phi_{(k)}^2 \alpha + V_{,(k)} \delta \phi_{(k)} \right), \quad
       \delta p = \sum_{k} \left( \dot \phi_{(k)} \delta \dot \phi_{(k)}
       - \dot \phi_{(k)}^2 \alpha - V_{,(k)} \delta \phi_{(k)} \right),
   \nonumber \\
   & & ( \mu + p ) v = {k \over a} \sum_{k} \dot \phi_{(k)} \delta \phi_{(k)},
       \quad
       \pi^{(s)} = 0.
   \label{MSFs-pert} 
\eea
In the mixture of the fluids and fields, the effective fluid quantities 
of the scalar fields in eqs. (\ref{MSFs-BG},\ref{MSFs-pert}) should be
added to the total fluid quantities in eq. (\ref{fluid-sum}).
Equation (\ref{EOM-MSFs}) gives
\bea
   & & \delta \ddot \phi_{(i)} + 3 H \delta \dot \phi_{(i)}
       + {k^2 \over a^2} \delta \phi_{(i)}
       + \sum_{j} V_{,(i)(j)} \delta \phi_{(j)}
       = \dot \phi_{(i)} \left( \kappa + \dot \alpha \right)
       + \left( 2 \ddot \phi_{(i)} + 3 H \dot \phi_{(i)} \right) \alpha.
   \label{EOM-MSFs-pert}
\eea
Equations (\ref{G1}-\ref{G5},\ref{MSFs-pert},\ref{EOM-MSFs-pert}) 
provide a complete set.
Under the gauge transformation we have
\bea
   & & \delta \tilde \phi_{(i)} = \delta \phi_{(i)} - \dot \phi_{(i)} \xi^t.
   \label{GT-2}
\eea
Thus, in a single component case, the uniform-field gauge 
$\delta \phi \equiv 0$ gives $v = 0$ which is the comoving gauge 
condition \cite{BST-1983}.

In order to match with eqs. (\ref{G6},\ref{G7}) it is convenient to
introduce fluid quantities of the individual field as
\bea
   & & \delta \mu_{(i)} \equiv \dot \phi_{(i)} \delta \dot \phi_{(i)}
       - \dot \phi_{(i)}^2 \alpha + V_{,(i)} \delta \phi_{(i)}, \quad
       \delta p_{(i)} \equiv \dot \phi_{(i)} \delta \dot \phi_{(i)}
       - \dot \phi_{(i)}^2 \alpha - V_{,(i)} \delta \phi_{(i)}, \quad
       v_{(i)} \equiv {k \over a} { \delta \phi_{(i)} \over \dot \phi_{(i)} },
       \quad
       \pi_{(i)}^{(s)} \equiv 0.
\eea
In this way, we can show $e_{(i)}$ in eq. (\ref{e-def}) becomes
\bea
   & & e_{(i)} = ( 1 - c_{(i)}^2 ) \delta \mu_{(i)v_{(i)}}.
   \label{e_i-MSFs}
\eea
With these definitions of fluid quantities
eqs. (\ref{G6},\ref{G7}) are satisfied with the following results
\bea
   & & \delta Q_{(i)} = \sum_{k} V_{,(i)(k)} \left(
       \dot \phi_{(k)} \delta \phi_{(i)} 
       - \dot \phi_{(i)} \delta \phi_{(k)} \right), \quad
       J_{(i)} = 0,
   \label{delta-Q-MSFs}
\eea
where we used eq. (\ref{EOM-MSFs-pert}) to derive eq. (\ref{G6}).
{}From eq. (\ref{e_i-MSFs}) we can derive \cite{Hwang-1991-PRW}
\bea
   & & e_{(ij)} = {1 \over 2} ( 2 - c_{(i)}^2 - c_{(j)}^2 )
       \left( S_{(ij)} + 3 H {a \over k} v_{(ij)} \right)
   \nonumber \\
   & & \qquad
       - ( c_{(i)}^2 - c_{(j)}^2 ) \Bigg\{ 
       {\delta \mu \over \mu + p} + 3 H {a \over k} v
       + {1 \over 2} \sum_k {\mu_{(k)} + p_{(k)} \over \mu + p}
       \left[ S_{(ik)} + S_{(jk)}
       + 3 H {a \over k} \left( v_{(ik)} + v_{(jk)} \right) \right] \Bigg\}.
   \label{e_ij-MSFs}
\eea
Thus, eq. (\ref{v_ij-eq}) simplifies to become
$\dot v_{(ij)} - 2 H v_{(ij)} = (k/a) S_{(ij)}$ which is satisfied identically;
these were found in \cite{Hwang-1991-PRW}.
Introducing a gauge-invariant combination
\bea
   & & \delta \phi_{(ij)} \equiv {\delta \phi_{(i)} \over \dot \phi_{(i)}}
       - {\delta \phi_{(j)} \over \dot \phi_{(j)}},
\eea
we have $v_{(ij)} = (k/a) \delta \phi_{(ij)}$,
and eq. (\ref{S_ij-eq}) leads to 
\bea
   & & \delta \ddot \phi_{(ij)} 
       - {3 \over 2} H \Bigg[ ( c_{(i)}^2 + c_{(j)}^2 ) \delta \dot \phi_{(ij)}
       + ( c_{(i)}^2 - c_{(j)}^2 ) \sum_k {\mu_{(k)} + p_{(k)} \over \mu + p}
       \left( \delta \dot \phi_{(ik)} + \delta \dot \phi_{(jk)} \right) \Bigg]
       + \left( - 3 \dot H + {k^2 \over a^2} \right) \delta \phi_{(ij)} 
       - \delta Q_{(ij)}
   \nonumber \\
   & & \qquad
       = 3 H (c_{(i)}^2 - c_{(j)}^2 ) {\delta \mu_v \over \mu + p},
   \label{delta-phi_ij-eq}
\eea
where, from eq. (\ref{delta-Q-MSFs}) we have
\bea
   & & \delta Q_{(ij)} = \sum_k \left(
       V_{,(i)(k)} {\dot \phi_{(k)} \over \dot \phi_{(i)}} \delta \phi_{(ik)}
       - V_{,(j)(k)} {\dot \phi_{(k)} \over \dot \phi_{(j)}} \delta \phi_{(jk)}
       \right).
   \label{Q_ij}
\eea
Using $c_{(i)}^2 = 1 + 2 V_{,(i)}/(3 H \dot \phi_{(i)})$ we have
\bea
   & & {1 \over a^3 \dot \phi_{(i)} \dot \phi_{(j)}} 
       \left( a^3 \dot \phi_{(i)} \dot \phi_{(j)} \delta \dot \phi_{(ij)}
       \right)^\cdot
       - \left( {V_{,(i)} \over \dot \phi_{(i)}}
       - {V_{,(j)} \over \dot \phi_{(j)}} \right) 
       \sum_k {\dot \phi_{(k)}^2 \over \mu + p}
       \left( \delta \dot \phi_{(ik)} + \delta \dot \phi_{(jk)} \right)
   \nonumber \\
   & & \qquad
       + \left( - 3 \dot H + {k^2 \over a^2} \right) \delta \phi_{(ij)}
       - \sum_k \left(
       V_{,(i)(k)} {\dot \phi_{(k)} \over \dot \phi_{(i)}} \delta \phi_{(ik)}
       - V_{,(j)(k)} {\dot \phi_{(k)} \over \dot \phi_{(j)}} \delta \phi_{(jk)}
       \right)
       = 2 \left( {V_{,(i)} \over \dot \phi_{(i)}}
       - {V_{,(j)} \over \dot \phi_{(j)}} \right) {\delta \mu_v \over \mu + p}.
   \label{delta-phi_ij-eq2}
\eea
Using eq. (\ref{Poisson-eq}), assuming $K = 0$,
eq. (\ref{delta-phi_ij-eq2}) becomes
eq. (26) in \cite{Hwang-Noh-2000-MSFs}; 
in the present case it is valid for general $K$.
{}From eqs. (\ref{e-def},\ref{MSFs-pert}) we have
$e = ( 1 - c_s^2 ) \delta \mu_v - 2 \sum_k V_{,(k)} \delta \phi_{(k)v}$.
Using eqs. (\ref{GT},\ref{GT-2}) we have
\bea
   & & e = ( 1 - c_s^2 ) \delta \mu_v
       - {2 \over \mu + p} \sum_{k,l} V_{,(k)} \dot \phi_{(k)}
       \dot \phi_{(l)}^2 \delta \phi_{(kl)}.
   \nonumber 
\eea
Thus, eq. (\ref{delta-mu_v}) gives
\bea
   & & \delta \mu_v = {k^2 - 3K \over 4 \pi G a^2} \varphi_\chi
       = - {1 - 3K/k^2 \over 1 - 3 ( 1 - c_s^2 ) K/k^2}
       \Big( {\mu + p \over H} \dot \Phi 
       - {2 \over \mu + p} \sum_{k,l} V_{,(k)} \dot \phi_{(k)} 
       \dot \phi_{(l)}^2 \delta \phi_{(kl)} \Big).
   \label{delta-mu_v-MSFs} 
\eea
The RHS of eq. (\ref{delta-phi_ij-eq2}) can be compared with the one of
eq. (\ref{S_ij-eq-2}), i.e., eq. (\ref{S_ij-eq-2-RHS}).
As mentioned below eq. (\ref{S_ij-eq-2-RHS}), in the large-scale limit
isocurvature modes decouple from the curvature one for
the multiple fluid system; 
in terms of $\varphi_\chi$ eq. (\ref{S_ij-eq-2-RHS}) shows $k^4/a^4$ factor. 
However, it is different for the multiple field system; 
in terms of $\varphi_\chi$ eq. (\ref{delta-phi_ij-eq2}) shows $k^4/a^4$ factor. 
Equation (\ref{Phi-eq}) remains valid with vanishing $\pi^{(s)}$, thus
\bea
   & & {H^2 c_A^2 \over ( \mu + p) a^3} \left[ { (\mu + p) a^3 \over 
       H^2 c_A^2} \dot \Phi \right]^\cdot + c_A^2 {k^2 \over a^2} \Phi
       = {2 H^2 c_A^2 \over (\mu + p) a^3}
       \left[ {a^3 \over (\mu + p) H c_A^2} \sum_{k,l}
       V_{,(k)} \dot \phi_{(k)} \dot \phi_{(l)}^2 \delta \phi_{(kl)}
       \right]^\cdot,
   \label{Phi-eq-MSFs}
\eea
where $c_A^2 \equiv 1 - 3 ( 1 - c_s^2 ) K/k^2$.
This equation can be compared with eq. (\ref{Phi-eq}).
Since $k^2 \rightarrow 0$ does not imply $c_A^2 k^2 \rightarrow 0$,
contrary to the ideal fluid situation, we have to be careful in examining
the large-scale limit for $K \neq 0$ case.
From eq. (\ref{delta_v-eq})
or from eq. (\ref{delta_v-multi-eq}), using eq. (\ref{e_i-MSFs}), we can derive
\bea
   & & {1 + w \over a^2 H} \left[ {H^2 \over a ( \mu + p)}
       \left( {a^3 \mu \over H} \delta_v \right)^\cdot \right]^\cdot
       + c_A^2 {k^2 \over a^2} \delta_v 
       = {k^2 - 3 K \over a^2} {1 \over \mu}
       {2 \over \mu + p} \sum_{k,l} V_{,(k)} \dot \phi_{(k)}
       \dot \phi_{(l)}^2 \delta \phi_{(kl)}. 
   \label{delta_v-eq-MSFs}
\eea

Equations (\ref{delta-phi_ij-eq2}-\ref{delta_v-eq-MSFs}) 
provide a complete set in terms of 
the curvature mode $\Phi$ (or $\delta_v$, or $\varphi_\chi$
using eq. (\ref{Poisson-eq})) and isocurvature modes $\delta \phi_{(ij)}$.
Various analytic solutions for the curvature and the isocurvature modes
in a system of scalar fields were studied 
in \cite{Hwang-Noh-2000-MSFs} assuming $K = 0 = \Lambda$.
Here we have considered general $K$ and $\Lambda$ 
in the background FLRW world model and presented the fluid formulation
of the system.
Literatures on the subject can be found in \cite{Hwang-Noh-2000-MSFs};
some selected ones are in \cite{MSFs}, and 
for recent additions, 
see \cite{Nibbelink-vanTent-2001,Bartolo-etal-2001,Starobinsky-etal-2001}.

%%%%%%%%%%%%%%%%%%%%%%%%%%%%%%%%%%%%%%%%%%%%%%%%%%%%%%%%%%%%%%
\subsection{Slow-roll: linear-order solutions}
                                                \label{sec:slow-roll}

We introduce the following slow-roll parameters 
(for different definitions, see 
\cite{Nibbelink-vanTent-2001,Bartolo-etal-2001})
\bea
   & & \epsilon \equiv {\dot H \over H^2}, \quad
       \epsilon_i \equiv {\ddot \phi_{(i)} \over H \dot \phi_{(i)}}, \quad
       \epsilon_{ij} \equiv {V_{,(i)(j)} \over 3 H^2}.
   \label{slow-roll}
\eea
In \S 4 of \cite{Hwang-Noh-2000-MSFs} we have presented various
solutions of eqs. (\ref{delta-phi_ij-eq2},\ref{Phi-eq-MSFs})
to the zero-th order slow-roll limit of $\epsilon_i$,
i.e., ignoring $\epsilon_i$, thus $V_{,(i)} = - 3 H \dot \phi_{(i)}$.
In this case eqs. (\ref{delta-phi_ij-eq2},\ref{Phi-eq-MSFs}) decouple 
from each other: see eqs. (29,30) in \cite{Hwang-Noh-2000-MSFs}. 
We consider $K = 0 = \Lambda$.
The assisted inflation based on multiple fields each with an
exponential potential \cite{assisted-inflation} has
$\epsilon_i = \epsilon = {\rm constant}$.
Perturbations in the assisted inflation can be managed in simple way
and were studied in \cite{assisted-inflation,assisted-pert}. 

In addition to zero-th order in $\epsilon_i$, by further assuming 
zero-th order in $\epsilon$, we have
\bea
   & & {1 \over a^3} \left( a^3 \dot \varphi_v \right)^\cdot
       + {k^2 \over a^2} \varphi_v = 0, \quad
       {1 \over a^3} \left( a^3 \delta \dot \phi_{(ij)} \right)^\cdot
       + {k^2 \over a^2} \delta \phi_{(ij)} = 0,
\eea
where we used that $V_{,(i)(k)}$ terms in eq. (\ref{delta-phi_ij-eq2})
are of the order $\epsilon$ and $\epsilon_i$.
Thus, the curvature and the isocurvature modes are decoupled, and in 
the large-scale limit we have the general solutions \cite{Hwang-Noh-2000-MSFs}
\bea
   & & \varphi_v (k,t) = C (k) - \tilde D (k) \int^t {1 \over a^3} dt, 
       \quad
       \delta \phi_{(ij)} (k,t) = C_{ij} (k) 
       - D_{ij} (k) \int^t {1 \over a^3} dt.
   \label{slow-roll-sol1}
\eea
Notice that the non-transient solutions remain constant in time.
{}From eqs. (\ref{GI},\ref{MSFs-pert}) we have
\bea
   & & \varphi_v 
       = - {H \over \mu + p} \sum_k \dot \phi_{(k)} \delta \phi_{(k)\varphi}.
\eea
In the single component scalar field, equation for $\delta \phi_\varphi$, 
which is $\delta \phi$ in the uniform-curvature gauge, 
resembles most closely the scalar field equation in the given 
(i.e., without accompanying metric fluctustions) curved background \cite{QFT}.
Equation (17) in \cite{Hwang-Noh-2000-MSFs} can be written using
the slow-roll parameters as
\bea
   & & \delta \ddot \phi_{(i)\varphi} + 3 H \delta \dot \phi_{(i)\varphi}
       + {k^2 \over a^2} \delta \phi_{(i)\varphi}
       = \sum_k \left[ 8 \pi G \dot \phi_{(i)} \dot \phi_{(k)}
       \left( 3 - \epsilon + \epsilon_i + \epsilon_k \right)
       - 3 H^2 \epsilon_{ik} \right] \delta \phi_{(k)\varphi},
   \label{delta-phi-i-eq-SR}
\eea
which is still exact.
Thus, notice that although equations for $\varphi_v$ and
$\delta \phi_{(ij)}$ are decoupled to the zero-th order
in the slow-roll parameters, equations for $\delta \phi_{(i)\varphi}$
are generally coupled to the same order.

Now, to the linear order in the slow-roll parameters,
and considering the non-transient solutions of eq. (\ref{slow-roll-sol1}),
eqs. (\ref{Phi-eq-MSFs},\ref{delta-phi_ij-eq2}) give
\bea
   & & {1 \over a^3} \left( a^3 \dot \varphi_v^{(1)} \right)^\cdot
       + {k^2 \over a^2} \varphi_v^{(1)} 
       = - {2 H^2 \over (\mu + p)^2} \sum_{j,k} 
       {1 \over a^3} \left( a^3 \epsilon_j \right)^\cdot
       \dot \phi_{(j)}^2 \dot \phi_{(k)}^2 \delta \phi_{(jk)}^{(0)}
       \equiv {\cal S}^{(1)},
   \label{cal-S} \\
   & & {1 \over a^3} \left( a^3 \delta \dot \phi_{(ij)}^{(1)} \right)^\cdot
       + {k^2 \over a^2} \delta \phi_{(ij)}^{(1)} 
       = 3 H^2 \left[ \epsilon \delta \phi_{(ij)}^{(0)}
       + \sum_k \left( \epsilon_{ik} {\dot \phi_{(k)} \over \dot \phi_{(i)}} 
       \delta \phi_{(ik)}^{(0)}
       - \epsilon_{jk} {\dot \phi_{(k)} \over \dot \phi_{(j)}} 
       \delta \phi_{(jk)}^{(0)} \right) \right]
       \equiv {\cal S}_{ij}^{(1)},
   \label{cal-S_ij}
\eea
where superscripts $(0)$ and $(1)$ indicate the variables to the
zero-th order and the linear order in the slow-roll parameters, respectively.
Apparently, we can extend this perturbative procedure to the higher order
in the slow-roll parameters.
Therefore, to the linear order in the slow-roll parameters 
the general solutions in the large-scale limit become
\bea
   & & \varphi_v^{(1)} = - \int^t a^3 
       \left( \int^t {1 \over a^3} dt \right) {\cal S}^{(1)} dt
       + \left( \int^t {1 \over a^3} dt \right) \int^t a^3
       {\cal S}^{(1)} dt, 
   \\
   & & \delta \phi_{(ij)}^{(1)} 
       = - \int^t a^3 
       \left( \int^t {1 \over a^3} dt \right) {\cal S}_{ij}^{(1)} dt
       + \left( \int^t {1 \over a^3} dt \right) \int^t a^3 
       {\cal S}_{ij}^{(1)} dt.
   \label{SR-sol2}
\eea
We note that to the linear order in the slow-roll parameter $\epsilon_i$
and considering the non-transient solution in the large-scale limit,
the RHS of eq. (\ref{delta-phi_ij-eq2}), using eq. (\ref{delta-mu_v-MSFs}),
vanishes.
Thus, in such a case the isocurvature modes decouple from the curvature one,
whereas the curvature mode is coupled with the isocurvature ones,
see eqs. (\ref{cal-S},\ref{cal-S_ij}).
Studies in more general situation with nonlinear sigma type coupling
are presented in \cite{Nibbelink-vanTent-2001} 
keeping covariant forms in field space.

%%%%%%%%%%%%%%%%%%%%%%%%%%%%%%%%%%%%%%%%%%%%%%%%%%%%%%%%%%%%%%
\section{Fluid and field system}
                                                \label{sec:fluid-field}

The formulations presented in \S \ref{sec:basic} and \ref{sec:Adiabatic} 
are valid for a mixture of arbitrary numbers of fluids and scalar fields.
Thus, we will not rewrite the complete set of equations again.

As an example, in this section we consider a two-component system of 
a fluid and a field with a general interaction term between them.
As the Lagrangian we consider
\bea
   & & {\cal L} = \sqrt{-g} \left[ {1 \over 16 \pi G} R 
       - {1 \over 2} \phi^{;c} \phi_{,c} - V(\phi) + L_{(f)} \right].
\eea
$L_{(f)}$ is the fluid part Lagrangian which also 
depends on the scalar field as 
$L_{(f)} = L_{(f)} ({\rm fluid}, g_{ab}, \phi)$.
The variation with respect to $g_{ab}$ leads to eq. (\ref{GFE}) with an 
identification $T_{ab} = T_{(f)ab} + T_{(\phi)ab}$.
The variation with respect to $\phi$ leads to the equation of motion for $\phi$
\bea
   & & \phi^{;c}_{\;\;\; c} - V_{,\phi} = - L_{(f),\phi} \equiv \Gamma.
   \label{EOM-ff}
\eea
Thus, from eq. (\ref{Tab-i-conservation}) we have
$Q_{(f)a} = - Q_{(\phi)a}$ and
\bea
   & & T_{(f)a;b}^{\;\;\;\; b} \equiv Q_{(f)a} = L_{(f),\phi} \phi_{,a}
\eea
where we used $i,j, \dots = 1,2 \equiv f,\phi$.
As we decompose $\Gamma = \bar \Gamma + \delta \Gamma$,
from eq. (\ref{Q-decomposition}) we have
\bea
   & & \bar Q_{(f)} = \bar \Gamma \dot {\bar \phi}, \quad
       \delta Q_{(f)} = \delta \Gamma \dot \phi + \Gamma ( \delta \dot \phi
       - \dot \phi \alpha ), \quad
       J_{(f)} = - \Gamma \delta \phi.
\eea
We have $\bar Q_{(\phi)} = - \bar Q_{(f)}$ and similarly for
$\delta Q_{(\phi)}$ and $J_{(\phi)}$.

By interpreting the fluid quantities properly, equations in 
\S \ref{sec:basic} and \ref{sec:Adiabatic} remain valid.
{}For the background we set 
$\mu = \mu_{(f)} + \mu_{(\phi)}$ and similarly for $p$ with
\bea
   & & \mu_{(\phi)} = {1 \over 2} \dot \phi^2 + V, \quad
       p_{(\phi)} = {1 \over 2} \dot \phi^2 - V.
\eea
Equation (\ref{BG1}) remains valid.
Equation (\ref{BG2}) is valid for $(i) = (f)$, and eq. (\ref{EOM-ff}) leads to
\bea
   & & \ddot \phi + 3 H \dot \phi + V_{,\phi} = - \Gamma.
   \label{EOM-ff-BG}
\eea
{}For the perturbations we set
$\delta \mu = \delta \mu_{(f)} + \delta \mu_{(\phi)}$ and similarly for
$\delta p$, $(\mu + p)v$, and $\pi^{(s)}$, with
\bea
   & & \delta \mu_{(\phi)} = \dot \phi \delta \dot \phi
       - \dot \phi^2 \alpha + V_{,\phi} \delta \phi, \quad
       \delta p_{(\phi)} = \dot \phi \delta \dot \phi
       - \dot \phi^2 \alpha - V_{,\phi} \delta \phi, \quad
       v_{(\phi)} = {k \over a} {\delta \phi \over \dot \phi}, \quad
       \pi^{(s)}_{(\phi)} = 0.
\eea
Equations (\ref{G1}-\ref{G5}) are valid for collective fluid quantities.
Equations (\ref{G6},\ref{G7}) are valid for the fluid with $(i) = (f)$,
and eq. (\ref{EOM-ff}) leads to
\bea
   & & \delta \ddot \phi + 3 H \delta \dot \phi + {k^2 \over a^2} \delta \phi
       + V_{,\phi\phi} \delta \phi
       = \dot \phi \left( \kappa + \dot \alpha \right)
       + \left( 2 \ddot \phi + 3 H \dot \phi \right) \alpha - \delta \Gamma.
   \label{G8-MSF}
\eea

After choosing the temporal gauge condition we can derive a fourth-order
differential equation, or a set of coupled two second-order differential
equations.
As an example, we derive one form of such a set of equations based on 
a gauge condition $v_{(f)}/k \equiv 0$ which is the comoving gauge based on 
the fluid velocity; in the following we sim ply write $v_{(f)} = 0$
as the gauge condition.
Under our gauge condition $v_{(f)} = 0$ the perturbation variables are gauge 
invariant, i.e., $\delta_{(f)} (\equiv \delta \mu_{(f)} / \mu_{(f)}) 
= \delta_{(f) v_{(f)}}$ and $\delta \phi = \delta \phi_{v_{(f)}}$,
where, from eqs. (\ref{GI},\ref{GT-2}), we have:
\bea
   & & \delta_{(f) v_{(f)}} \equiv \delta_{(f)} 
       - {\dot \mu_{(f)} \over \mu_{(f)}} {a \over k} v_{(f)}, \quad
       \delta \phi_{v_{(f)}} \equiv \delta \phi 
       - \dot \phi {a \over k} v_{(f)}.
\eea
We {\it assume} an ideal fluid with $w \equiv p_{(f)}/\mu_{(f)} 
= {\rm constant}$.
Using eq. (\ref{G7}) for the fluid component we can express $\alpha$ in 
terms of $\delta_{(f)}$ and $\delta \phi$.
Using eq. (\ref{G6}) we can express $\kappa$ in terms of 
$\delta_{(f)}$ and $\delta \phi$.
Thus, eqs. (\ref{G5},\ref{G8-MSF}) lead to
\bea
   & & \ddot \delta_{(f)} + \left( 2 - 3 w \right) H \dot \delta_{(f)} 
       + \Bigg[ w {k^2 \over a^2} 
       - 6 w \left( \dot H + H^2 \right) 
       - 4 \pi G \mu_{(f)} (1 + w) (1 + 3 w) \Bigg] \delta_{(f)}
   \nonumber \\
   & & \qquad
       = 8 \pi G \left[ 2 w \dot \phi^2 \delta_{(f)}
       + (1 + w) \left( 2 \dot \phi \delta \dot \phi 
       - V_{,\phi} \delta \phi \right) \right]
   \nonumber \\
   & & \qquad
       - {1 \over a^2} \left\{ a^2 \left[ {\Gamma \dot \phi \delta_{(f)}
       \over \mu_{(f)}} 
       - {\dot \phi \delta \Gamma \over \mu_{(f)}} 
       - {\Gamma \over \mu_{(f)}} \left( \delta \dot \phi + 3 H \delta \phi
       \right) \right] \right\}^\cdot
       + \left( 3 \dot H + 16 \pi G \dot \phi^2
       - {k^2 \over a^2} \right) {\Gamma \over \mu_{(f)}} \delta \phi,
   \label{eq1} \\
   & & \delta \ddot \phi + 3 H \delta \dot \phi 
       + \left( {k^2 \over a^2} + V_{,\phi\phi} \right) \delta \phi
       = {1 \over 1 + w} \left[ (1- w) \dot \phi \dot \delta_{(f)}
       - 2 w \left( \ddot \phi + 3 H \dot \phi \right) 
       \delta_{(f)} \right]
       + {\Gamma \dot \phi^2 \over (1 + w) \mu_{(f)}} \delta_{(f)}
   \nonumber \\
   & & \qquad
       - \left[ 1 + {\dot \phi^2 \over (1 + w) \mu_{(f)}} \right] 
       \delta \Gamma
       + {1 \over 1 + w} \Bigg[ - {\Gamma \dot \phi \over \mu_{(f)}} 
       \delta \dot \phi
       - \dot \phi \left( {\Gamma \delta \phi \over \mu_{(f)}} \right)^\cdot
       + 2 \left( \ddot \phi + 3 H \dot \phi \right) 
       {\Gamma \delta \phi \over \mu_{(f)}} \Bigg].
   \label{eq2}
\eea
Although the RHS of eq. (\ref{eq1})
contains $\delta \ddot \phi$ term, it can be replaced by using 
eq. (\ref{eq2}).  
Notice that eqs. (\ref{eq1},\ref{eq2}) are valid for general $K$ and $\Lambda$.
The LHS of eq. (\ref{eq1}) is the familiar density perturbation
in the comoving gauge derived in \cite{Nariai-1969,Bardeen-1980},
and the LHS of of eq. (\ref{eq2}) is also the familiar perturbed scalar field
equation without the metric perturbation; compare with eq. (\ref{G8-MSF}).
Thus, our $v_{(f)} = 0$ gauge choice allows simple equations for the
uncoupled parts of the system. 

This set of equations or eqs. (\ref{delta_v-multi-eq},\ref{e_i-MSFs}) 
will be useful to handle the following situations:

\noindent
(i) Warm inflation scenario with nonvanishing 
interaction term, $\Gamma$ and $\delta \Gamma$, between the field and 
the radiation ($w = {1 \over 3}$) \cite{warm-inflation}.

\noindent
(ii) Time-varying cosmological constant simulated using the scalar field,
often called a quintessence \cite{quintessence};
in this case we may ignore the direct interaction between the fluid 
and the field $\Gamma = 0 = \delta \Gamma$.
An application of eqs. (\ref{eq1},\ref{eq2}) to a system
with an exponential type field potential is made in \cite{Hwang-Noh-2001-Q}.

%%%%%%%%%%%%%%%%%%%%%%%%%%%%%%%%%%%%%%%%%%%%%%%%%%%%%%%%%%%%%%
\section{Summary}
                                                \label{sec:Summary}

In this paper we have investigated aspects of scalar-type cosmological 
perturbation in the context of multiple numbers of mutually interacting 
imperfect fluids and minimally coupled scalar fields in Einstein gravity.
Equations are presented using the curvature ($\Phi$ or $\varphi_\chi$)
and isocurvature ($S_{(ij)}$ or $\delta \phi_{(ij)}$) perturbation variables.
It looks there exists no clear consensus about the curvature/isocurvature
decomposition of perturbation in the literature.
Since the equations are all coupled, such a decomposition may have clearer
meaning in setting up the initial conditions.
In \S \ref{sec:Adiabatic} and \ref{sec:MSFs} we have shown that
the equations can be meaningfully classified into the our definitions
of the curvature ($\Phi$ or $\varphi_\chi$) and isocurvature ($S_{(ij)}$ 
or $\delta \phi_{(ij)}$) modes; see below eq. (\ref{delta-mu_v}).
We have shown that either $\Phi$ or $\varphi_\chi$ can 
characterize the curvature mode depending on situations.

We have presented the equations in a gauge-ready form where we have a
freedom to choose one temporal gauge condition from the variables in
eqs. (\ref{GT},\ref{GT-2}) or the linear combinations of them. 
In the general situation, eqs. (\ref{G1}-\ref{G9}) provide a complete set.
In the case of scalar fields we also have 
eqs. (\ref{EOM-MSFs-pert},\ref{MSFs-pert}). 
In terms of the curvature and isocurvature perturbations,
eqs. (\ref{Phi-eq},\ref{S_ij-eq},\ref{v_ij-eq},\ref{e-multi},\ref{delta-mu_v})
with eqs. (\ref{G1}-\ref{G5}) provide a complete set, 
and in the special case with multiple fields 
eqs. (\ref{delta-phi_ij-eq}-\ref{Phi-eq-MSFs}) 
or eqs. (\ref{delta_v-multi-eq},\ref{e_i-MSFs}) provide complete sets.
In \S \ref{sec:fluid-field} we have shown how to consider
the case with mutual interaction among fluids and fields.

Some new results found in this paper are the following:

\noindent
(i)   Gauge-ready formulation of the fluid-field system in Einstein gravity,
      \S \ref{sec:basic}, \ref{sec:MSFs} and \ref{sec:fluid-field}.

\noindent
(ii)  Several useful forms of the individual fluid equations,
      eqs. (\ref{delta-i_CG-eq},\ref{v-i_ZSG-eq}) and \S \ref{sec:component}.

\noindent
(iii) Analyses of various curvature perturbations in realistic
      situations of multi-component system, \S \ref{sec:Curvature-pert}.

\noindent
(iv) Equations (\ref{S_ij-eq},\ref{v_ij-eq}) are valid even in 
      a class of generalized gravity theories, see below
      eq. (\ref{G7}) and eq. (\ref{e_ij-def}).

\noindent
(v)  Extension of fluid formulation of the multiple field system,
      \S \ref{sec:MSFs}.

\noindent
(vi)   In the scalar field system the isocurvature modes
      are less decoupled from the curvature mode in the large-scale limit
      compared with the fluids system,
      see below eq. (\ref{delta-mu_v-MSFs}). 
      The couplig term vanishes to the linear order in slow-roll expansion,
      see below eq. (\ref{SR-sol2}).

\noindent
(vii) Solutions for $c_{(i)}^2 = c_{(j)}^2$ fluids, 
      eqs. (\ref{S_ij-eq-3}-\ref{S_ij-eq-3-sol}).

\noindent
(viii)  Solutions valid to the linear order in the slow-roll parameters,
      \S \ref{sec:slow-roll}.

The sets of equations in \S \ref{sec:fluid-field} may be useful to analyze 
the evolution of structures in the world models with quintessence
and in the warm inflation scenario.
However, these sets of equations are the ones derived in certain gauge
conditions, and when we encounter new problems we believe it is best to go
back to the original set of equations in the gauge-ready form and see whether
we could gather any new perspective from other gauge conditions as well:
the set is in eqs. (\ref{G1}-\ref{G9},\ref{EOM-MSFs-pert}).
Corresponding set of equations in a gauge-ready form 
which is applicable in a wide class of generalized gravity theories 
including contributions from the kinetic components based on
Boltzmann equations is presented in \cite{Hwang-1991-PRW,Hwang-Noh-2001-CMBR}.
Specific applications will be made in future occasions.

%%%%%%%%%%%%%%%%%%%%%%%%%%%%%%%%%%%%%%%%%%%%%%%%%%%%%%%%%%%%%%
\subsection*{Acknowledgments}

We thank Winfried Zimdahl for suggesting the study and
for his useful discussions and comments during the work.
We also wish to thank Stefan Groot Nibbelink and David Wands for useful 
comments.
HN was supported by grant No. 2000-0-113-001-3 from the
Basic Research Program of the Korea Science and Engineering Foundation.
JH was supported by the Korea Research Foundation Grants 
(KRF-2000-013-DA004 and 2000-015-DP0080).

\baselineskip 0pt
%%%%%%%%%%%%%%%%%%%%%%%%%%%%%%%%%%%%%%%%%%%%%%%%%%%%%%%%%%%%%

%%%%%%%%%%%%%%%%%%%%%%%%%%%%%%%%%%%%%%%%%%%%%%%%%%%%%%%%%%%%%%
\end{document}